\documentclass{aa}

\usepackage{graphicx}
\usepackage{amsmath}
\usepackage{amssymb}
\usepackage{tabularx}
\usepackage{booktabs}
\usepackage{siunitx}
\usepackage{array,multirow}
\usepackage[normalem]{ulem}
\usepackage{txfonts}
\usepackage[colorlinks,breaklinks]{hyperref}
\hypersetup{linkcolor=blue,citecolor=blue,filecolor=black,urlcolor=blue}
\usepackage{xspace}
\newcommand{\pkg}[1]{\textsc{\texttt{#1}}}
\newcommand{\hypergal}{\pkg{HyperGal}\xspace}
\newcommand{\sref}{\ensuremath{\text{ref}}}

\usepackage[switch]{lineno}
\usepackage{etoolbox} 
\newcommand*\linenomathpatch[1]{%
  \cspreto{#1}{\linenomath}%
  \cspreto{#1*}{\linenomath}%
  \csappto{end#1}{\endlinenomath}%
  \csappto{end#1*}{\endlinenomath}%
}
\linenomathpatch{equation}
\linenomathpatch{gather}
\linenomathpatch{multline}
\linenomathpatch{align}
\linenomathpatch{alignat}
\linenomathpatch{flalign}

\begin{document}
\title{HyperGal: hyperspectral scene modeling for supernova typing with
  the Integral Field Spectrograph SEDmachine}

\titlerunning{Hypergal}
\authorrunning{J.~Lezmy et al.}

\author{%
  J. Lezmy \inst{1},
  Y. Copin\inst{1},
  M. Rigault\inst{1},
  M. Smith\inst{1},
  J. D. Neill\inst{2}}

\institute{%
  Université de Lyon, Université Claude Bernard Lyon 1, CNRS/IN2P3, IP2I
  Lyon, F-69622, Villeurbanne, France\\
  \email{lezmy@ip2i.in2p3.fr} %
  \and %
  Division of Physics, Mathematics, and Astronomy, California Institute
  of Technology, Pasadena, CA 91125, USA
}

\date{Received \today; accepted xxx xxx xxx}

\abstract
{
  Recent developments in time domain astronomy, like the Zwicky
  Transient Facility, have made possible a daily scan of the entire
  visible sky, leading to the discovery of hundreds of new transients
  every night.  Among these detections, 10 to 15 are supernovae (SNe),
  which have to be classified prior to cosmological use.  The Spectral
  Energy Distribution machine (SEDm), a low resolution
  ($\mathcal{R} \sim 100$) Integral Field Spectrograph, has been
  designed, built, and operated to spectroscopically observe and
  classify targets detected by the ZTF main camera.  }
{
  As the current \pkg{pysedm} pipeline can only handle isolated point
  sources, it is limited by contamination when the transient is too
  close to its host galaxy core; this can lead to an incorrect typing
  and ultimately bias the cosmological analyses, and affect the SN
  sample homogeneity in terms of local environment properties.  We
  present a new scene modeler to extract the transient spectrum from
  its structured background, aiming at improving the typing efficiency
  of the SEDm.}
{
  \hypergal is a fully chromatic scene modeler, which uses archival
  pre-transient photometric images of the SN environment to generate a
  hyperspectral model of the host galaxy; it is based on the
  \pkg{cigale} SED fitter used as a physically-motivated spectral
  interpolator.  The galaxy model, complemented by a point source for
  the transient and a diffuse background component, is projected onto
  the SEDm spectro-spatial observation space and adjusted to
  observations; the SN spectrum is ultimately extracted from this
  multi-component model.  The full procedure, from scene modeling to
  transient spectrum extraction and typing, is validated on 5000
  simulated cubes built from actual SEDm observations of isolated host
  galaxies, covering a large variety of observing conditions and scene
  parameters.}
{
  We introduce the contrast $c$ as the transient-to-total flux ratio
  at SN location, integrated over the ZTF $r$ band.  From estimated
  contrast distribution of real SEDm observations, we show that
  \hypergal correctly classifies $\sim 95\%$ of SNe~Ia, and up to 99\%
  for contrast $c \gtrsim 0.2$, representing more than 90\% of the
  observations.  Compared to the standard point-source extraction
  method (without the hyperspectral galaxy modeling step), \hypergal
  correctly classifies 20\% more SNe~Ia between $0.1 < c < 0.6$ (50\%
  of the observation conditions), with less than 5\% of SN~Ia
  misidentifications.  The false positive rate is less than 2\% for
  $c > 0.1$ ($> 99\%$ of the observations), which represents half as
  much as the standard extraction method.  Assuming a similar contrast
  distribution for core-collapse SNe, \hypergal classifies 14\%
  additional SNe~II and 11\% additional SNe~Ibc.}
{
  \hypergal proves to be extremely effective to extract and classify
  SNe in the presence of strong contamination by the host galaxy,
  providing a significant improvement with respect to the single point
  source extraction.}


\maketitle
%


\section{Introduction}

In the last two decades, time-domain astronomy has become increasingly
efficient, thanks to the ability of the surveys to (near) daily scan
the entire visible sky.  We can cite the Catalina Real-Time Transient
Survey \citep{catalina}, PanSTARRS-1 \citep{panstarrs}, ASAS-SN
\citep{Asas} and ATLAS \citep{Atlas}.  A more recent survey is the
Zwicky Transient Facility \citep[ZTF,][]{ztfbellm, ztfgraham},
successor of the Palomar Transient Facility \citep{PTF}, and using a
\SI{47}{deg^{2}} camera.  With such equipment, ZTF detects
$\mathcal{O}(10^2)$ transients of interest every night, instrumental
artifacts and previously known sources excluded, with a typical
$5\sigma$ $r$ band AB magnitude limit of $20.5$. Among them, 10 to 15 are
new objects that have just appeared and became bright enough to be
detected.  Once the photometric detection is triggered, ZTF relays the
alert to the Spectral Energy Distribution machine
\citep[SEDm,][]{SEDm}, an Integral Field Spectrograph (IFS) designed
and built to spectroscopically type transients brighter than
$\sim 19.5$~mag, and operating on the Palomar 60-inch telescope.  The
core of the SEDm is a Micro-Lenslet Array (MLA) covering
$28\arcsec\times 28\arcsec$, subdivided into $52 \times 45$ hexagonal
spaxels, combined to a multi-band ($ugri$) field acquisition camera,
used for positioning and guiding.

Currently, the automated pipeline routinely used for IFS data reduction and
supernova (SN) spectrum extraction is \pkg{pysedm} \citep{pySEDm}.  Since this
pipeline intrinsically assumes the target is an isolated point source, it
cannot properly handle the situation where the transient is close to its host
galaxy core.

As a matter of fact, since August 2018, $\sim 30\%$ of the observed SN
show some severe host contamination which significantly decreases the
confidence level of the classification, and $\sim 10\%$ are just unusable.
This situation has various undesirable effects.  From a mere statistical point
of view, discarding SNe with too strong a host contamination reduces the
type~Ia SN (SN~Ia) sample by 10 to 20\%, which weakens the strength of the
Hubble diagram anchor at low redshift.  Furthermore, the wrong classification
of SNe~Ia could induce a significant bias in the cosmological analysis
\citep[e.g.][]{jonesSNcontam}.

Finally, a more subtle effect is related to the galactic environment bias,
which would be caused by selecting out host-contaminated SNe
\citep{Rigault2013}.  In the past years, numerous studies have shown that the
SN~Ia standardized luminosity is tightly correlated with the environment
properties.  \citet{rigault15, lssf_rigault20} showed that, after
  standardization for light curve shape and color, SNe~Ia having a large local
specific Star Formation Rate
are fainter by $0.16\pm 0.03$~mag.  Other tracers, like host galaxy stellar
mass \citep{kelly10, sullivan10, childress13, Betoule2014} or just host
morphology \citep{morph20}, are finding the same correlation between SN~Ia
luminosity and their environment.  Recently, \citet{briday} have shown that all
these tracers are compatible with two SN~Ia populations differing in
standardized magnitude by at least $0.12 \pm 0.01$~mag.

Some developments have been made to improve the robustness of the point
  source extraction by estimating the faintest iso-magnitude contour separating
  the galaxy and the SN \citep{contsep}; however, this is not yet optimal in
  most problematic situations, i.e. when the SN is faint or located near the
  host core: it only brings a marginal $1.7\%$ improvement in classification
  accuracy from the standard \pkg{pysedm} analysis.

One could think of handling the host contamination by interpolating the
  galaxy area under the transient from the external parts in the FoV.
  Unfortunately, there are several reasons for not using such method, beyond
  the mere signal-to-noise issue.  First, the seeing, which makes the SN spread
  over the galaxy structure: as much as the host light is contaminating the SN
  flux, the reverse is also true, and it is not clear how far from the SN
  position one could consider the galaxy flux to be free of the point source
  signal.  Furthermore, the host spatial structure under the SN extent --
  linear, concave or convex -- is not known \textit{a priori}, specially in a
  strongly structured region such as the galaxy core, which would prevent a
  clean and robust interpolation.  Finally, an interpolation would assume that
  the host spectral features are spatially uniform under the SN extent, which
  again is usually not the case, specially close to the galaxy core.

In order to improve the final SN~Ia sample in numerous ways, we present in this
paper \hypergal\footnote{The code is available online at
  {\url{https://github.com/JeremyLezmy/HyperGal}}.}, a scene modeler
specifically designed to handle the strong host contamination case, through a
detailed hyperspectral galaxy modeling, complemented by a smooth background
component and a point-source transient.  The algorithm concept is based on two
ideas: first, public multi-band wide photometric surveys can provide reference
information on the host galaxy before the transient event; second, the required
host galaxy cube (two spatial dimensions and one spectral one) can be estimated
from pure photometric observations using a dedicated SED fitter as a physically
motivated spectral interpolator.  The resulting hyperspectral host model can
then be projected in the observable space of the SEDm, taking into account all
observational effects: relative geometry between the photometric pixels (px)
and the IFS spaxels (spx), spatial (Point Spread Function, PSF) and spectral
(Line Spread Function, LSF) Impulse Response Functions (IRF) of the SEDm,
Atmospheric Differential Refraction (ADR), sky background and additional
diffused light.

Sec.~\ref{sec:pipeline} describes the \hypergal pipeline, and
validation tests on realistic simulations are presented in
Sec.~\ref{sec:validation}, to estimate the accuracy of the SN
extraction as well as the SN typing itself, since this is what the
SEDm is designed for.  We will also show the improvement with respect
to an isolated source extractor such as \pkg{pysedm}.  A discussion of
some hypotheses and possible future improvements can be found in
Sec.~\ref{sec:discuss}.

\section{\hypergal pipeline}
\label{sec:pipeline}

\begin{figure*}
  \centering
  \includegraphics[width=\textwidth]{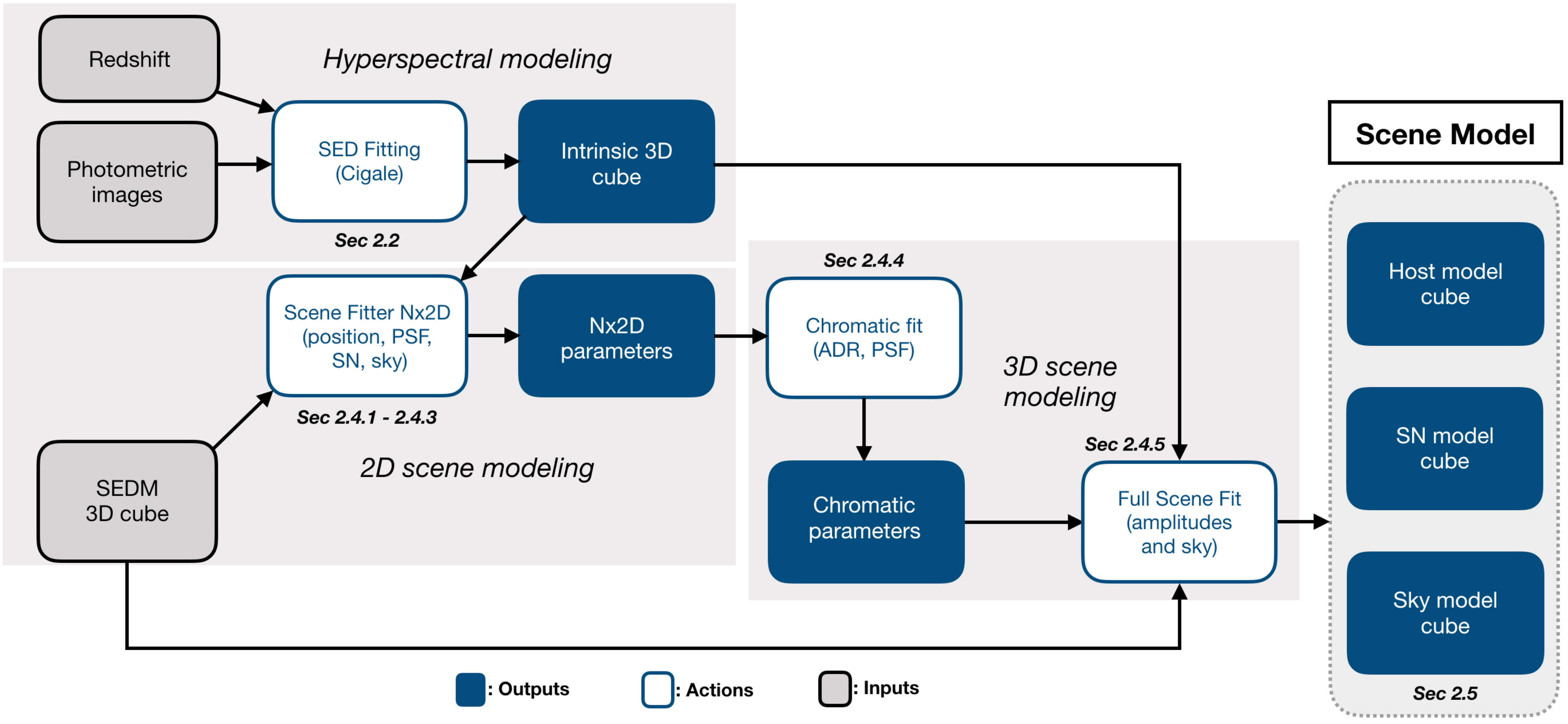}
  \caption{Main processing steps of the \hypergal pipeline, and the
    sections where they are detailed.}
  \label{fig:dag}
\end{figure*}

This section presents the different processing steps from the required
input to the transient spectrum extraction (Fig.~\ref{fig:dag}).  SN
ZTF20aamifit, at redshift of $z=0.045$ as measured from strong
H$\alpha$ line in the host spectrum, will systematically be used for
illustration; it was observed with the SEDm in February 17, 2020, at
airmass~1.7 in poor seeing conditions ($2\farcs 4$ FWHM).  The SN
is $\sim2\farcs8$ away from its host galaxy core, close enough not to
be considered as isolated (see Fig.~\ref{fig:cutout}).

\subsection{Inputs}
\label{sec:inputs}

Three main inputs are necessary to \hypergal: the SEDm cube to be
analysed, the archival photometric thumbnails, and the redshift of the
target.

\begin{figure}
  \centering
  \includegraphics[width=\linewidth]{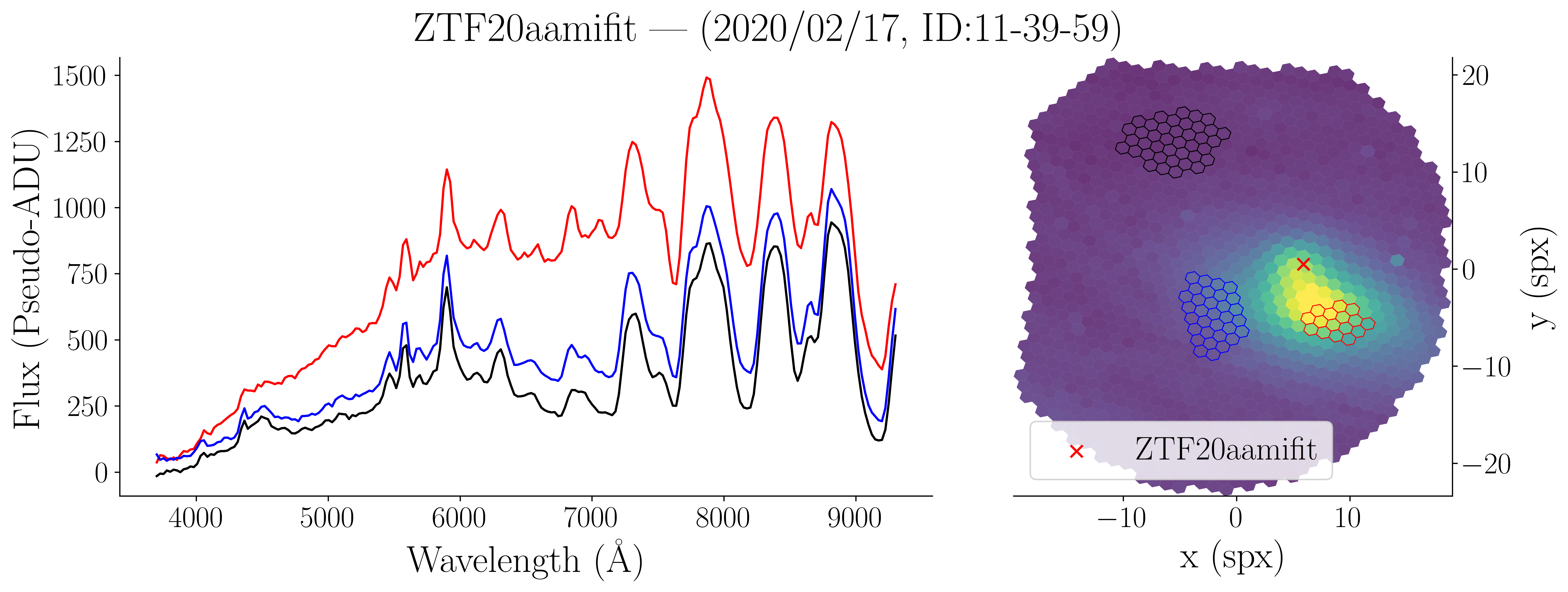}
  \caption{SEDm cube from the observation of ZTF20aamifit. The
    \emph{left panel} shows the spectra, whose color corresponds to
    the selected spaxels in the \emph{right panel} (white image of the
    spectrally integrated cube). The \emph{red cross} shows the SN
    position.}
    \label{fig:e3dsedm}
\end{figure}

The SEDm IFS $(x, y, \lambda)$ cube of the scene is built from the 2D
raw spectroscopic exposures with \pkg{pysedm} \citep[Sec.~2]{pySEDm}.
It includes all the components -- transient point source, spatially
and spectrally structured host galaxy, night sky background and
spatially smooth diffused light -- to be handled by the scene modeler
(Fig.~\ref{fig:e3dsedm}).

\begin{figure}
  \centering
  \includegraphics[width=\linewidth]{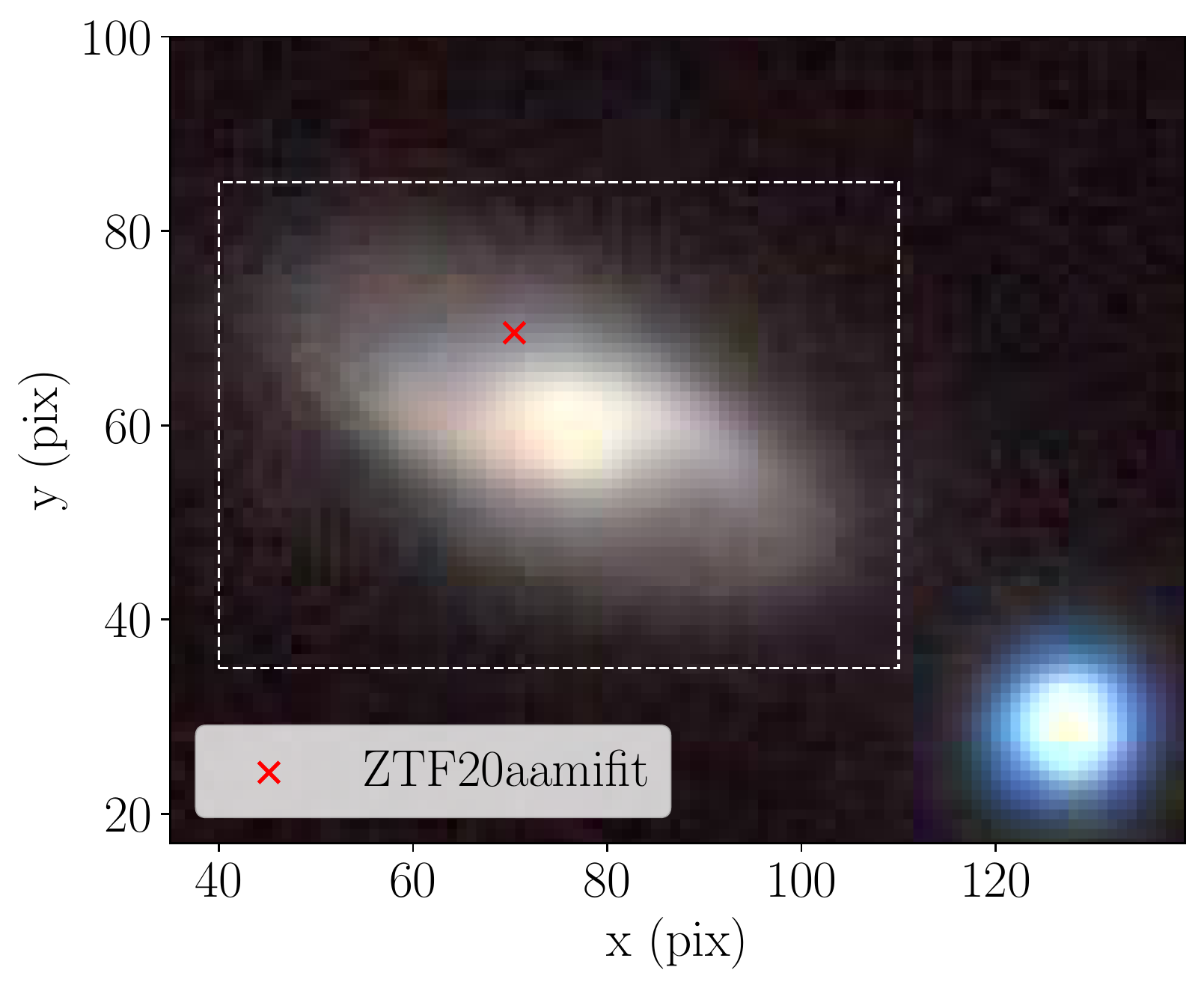}
  \caption{RGB image of the host galaxy of SN ZTF20aamifit,
    constructed from the PS1 $grz$ cutouts. The red cross shows the
    position of the SN detected by ZTF. The $x$- and $y$-axes are in
    native PS1 pixels, $0\farcs 25$ aside. The white dashed box will
    be used as boundaries in Fig.~\ref{fig:cig_spatial_check}
    and~\ref{fig:cig_spectral_check}.}
  \label{fig:cutout}
\end{figure}

The archival multi-band photometric images of the transient
environment, acquired \emph{before} the SN explosion, are obtained
from the PanSTARRS-1 (PS1) $3\pi$ Steradian survey
\citep{ChambersPanstarrs} in all $grizy$ bands, and queried at the SN
location through the Image Cutout Server%
\footnote{\url{https://ps1images.stsci.edu/cgi-bin/ps1cutouts}}.  PS1
is chosen for its sky coverage compatible with ZTF (north of
declination \SI{-30}{deg}).  Figure~\ref{fig:cutout} shows an RGB
image for ZTF20aamifit host galaxy, through the PS1 $grz$ bands.

An analysis of spatially structured scenes (harboring 3 or more well resolved
objects in the SEDm FoV) provides a precise estimation of a scale ratio of SEDm
and PS1 pixel sizes of $2.230 \pm 0.003$, which, for a PS1 px scale of
$0\farcs 25$, corresponds to an effective SEDm spaxel size of $0\farcs 558$.
Once measured, this SEDm scale is fixed in the pipeline. To save
computation time for the SED fit and the spatial projection step, PS1 images
are first spatially rebinned $2\times2$.

The third input is the host galaxy redshift, required by the SED-based
interpolation of the photometric images.  Around 50\% of the targets
observed by the SEDm have a host galaxy spectroscopic redshift known
beforehand \citep{ztfspecred}; for the others, a redshift is a
priori estimated from a preliminary transient spectrum extraction, using
the transient spectral features and the possible presence of emission
lines from the host galaxy. While it would be theoretically possible to
assess the host redshift directly during the scene modeling, we did not
try to implement this feature yet (see
Sec~\ref{sec:discuss}). Furthermore, the consequence of an inaccurate
input redshift has not been studied for this analysis.

\subsection{SED fit}
\label{sec:sedfit}

The SED fit aims to generate an effective hyperspectral -- i.e. full
3D $(x,y,\lambda)$ -- host model from the $grizy$ PS1 broadband
images.  During the process, each photometric pixel is treated
independently, so that the resulting spaxel in the output cube gets
its own spectrum.  At the end of this process, this cube is still
independent of the SEDm observation details (impulse responses,
atmospheric effects, etc.).  It is important to note that the SED
fitter is not used here to derive accurate and spatially resolved
physical parameters from the host galaxy, but rather to build a
physically plausible spectral interpolation compatible with broadband
archival images.

The software used for this step is \pkg{cigale}\footnote{Version 2020,
  \href{https://cigale.lam.fr}{https://cigale.lam.fr}}
\citep{Burgarella2005, Noll2009, cigale}.  It is based on a
progressive computation, successively using modules describing a
unique component of the SED.  The set of all parameters tested by
\pkg{cigale} is shown in Table~\ref{tab:cigaleparams}.

\subsubsection{Star Formation History and population}

The time-evolution of the Star Formation Rate (SFR) is described by
the Star Formation History (SFH) through the \texttt{sfhdelayed}
module.  Our SFH scenario includes two components, a delayed SFR and a
late burst:
\begin{equation}
  \label{eq:sfhdelayed}
  \text{SFR}(t) = \text{SFR}_{\text{delayed}}(t) + \text{SFR}_{\text{burst}}(t).
\end{equation}
Both terms have a decreasing exponential form,
\begin{align}
  \label{eq:sfh}
  \text{SFR}_{\text{delayed}}(t)
  &\propto \left(t/\tau_{\text{main}}^{2}\right) e^{-t/\tau_{\text{main}}} \\
  \text{SFR}_{\text{burst}}(t)
  &\propto e^{-(t-t_{0})/\tau_{\text{burst}}}
    \quad\text{for}\; t > t_{0},\;0\;\text{otherwise}.
\end{align}
The amplitude of the late starburst is fixed by the parameter
$f_{\text{burst}}$, defined as the ratio between the stellar mass
formed during this event and the total stellar mass.  The SFH is
applied with the Initial Mass Function (IMF) from \citet{Chabrier2003}
on the stellar population model from \citet{bc03}, used through the
\texttt{bc03} module.

\subsubsection{Nebular emission}

The light emitted in the Lyman continuum by the heaviest stars ionizes
the gas in the galaxy.  This physical process generates significant
radiative emission in the continuum and spectral lines.  This SED
component is described by the \texttt{nebular} module, based on
\citet{Inoue2011}.  The model is effectively parameterized by the
metallicity $Z$ (the same as in the stellar population model
\texttt{bc03}) and the ionization parameter $\log(U)$.

\subsubsection{Dust extinction}

Dust in the galaxy absorbs the radiation at short wavelengths,
especially from the UV to the near-IR; this energy is then re-emitted
in the mid- to far IR.  As \hypergal is primarily targeting sources at
redshift $z < 0.1$ in the optical domain, extinction effect is
properly considered through the dust attenuation module
\texttt{dustatt\_modified\_CF00} from \citet{dustatt}.  This approach
is considering two star populations: the young ones
($< \SI{e7}{years}$) still reside in their Birth Cloud (BC), and the
old ones are considered as already dispersed in the InterStellar
Medium (ISM).  Attenuation is therefore treated differently: for the
young population, both ISM and BC are considered, while for the old
population, only the ISM is considered.  In both case, the attenuation
$A_{\lambda}$ is modeled by a power law, normalized by the $V$-band
attenuation:
\begin{equation}
  A_{\lambda}^{k} = A_{V}^{k}\left(\frac{\lambda}{\lambda_V}\right)^{n_{k}}
  \quad k=\text{BC or ISM},
\end{equation}
with $\lambda_{V} = \SI{0.5}{\micro\meter}$.  The young-to-old star
$V$-band attenuation ratio is parameterized through
$\mu = A_{V}^{\text{ISM}}/(A_{V}^{\text{ISM}} + A_{V}^{\text{BC}})$, a
free parameter allowing more flexibility and a better estimate of the
H$\alpha$ emission lines \citep{Battisti2016, Buat2018, Malek2018,
  Chevallard2019}.  The power-law slope for the ISM is fixed at
$n_{\text{ISM}} = -0.7$ following \citet{dustatt}, and the slope for
the BC at $n_{\text{BC}} = -1.3$ as advocated in
\citet{Cunha2008MAGPHYS}.

For completeness, the \texttt{dale2014} module is used for the dust
emission \citep{dale2014}; however, this complex component has no
significant impact in our spectral domain.

\begin{table*}
  \centering
  \caption{Modules and input parameters used with \pkg{cigale}.}
  \label{tab:cigaleparams}
  \begin{tabular}{lcc}
    \toprule
    \textbf{Parameters} & \textbf{Symbol} & \textbf{Tested values}\\
    \midrule
    \multicolumn{3}{l}{\textbf{Star Formation History (SFH)}} \\
    e-folding time of the main stellar population
                        & $\tau_{\text{main}}$ (Gyr) & 1, 3, 5 \\
    e-folding time of the late starburst population
                        & $\tau_{\text{burst}}$ (Gyr) & 10 \\
    age of the main stellar population
                        & $\text{age}_{\text{main}}$ (Gyr) & 1, 2, 4, 8, 10, 12 \\
    age of the late starburst
                        & $\text{age}_{\text{burst}}$ (Myr) & 10, 40, 70 \\
    mass fraction of the late starburst
                        & $f_{\text{burst}}$
                        & 0, \num{e-3}, \num{e-2}, \num{e-1}, \num{2e-1} \\
    \hline
    \multicolumn{3}{l}{\textbf{Stellar population} } \\ 
    Metallicity & $Z$ & \num{e-4}, \num{4e-4}, \num{4e-3}, \\
                        && \num{8e-3}, \num{2e-3}, \num{5e-2} \\
    \hline
    \multicolumn{3}{l}{\textbf{Nebular emission}} \\
    Ionisation parameter& $\log(U)$& $-4$, $-3$, $-2$, $-1$ \\
    \hline
    \multicolumn{3}{l}{\textbf{Dust attenuation}} \\ 
    InterStellar Medium attenuation in $V$
                        & $A_{V}^{\text{ISM}}$ & 0, 0.3, 0.7, 1, 1.3, 1.7, 2 \\
    $A_{V}^{\text{ISM}}/(A_{V}^{\text{ISM}}+A_{V}^{\text{BC}})$
                        & $\mu$ & 0.1, 0.3, 0.7, 1 \\
    BC power-law slope  & $n_{\text{BC}}$ & $-1.3$ \\
    ISM power-law slope & $n_{\text{ISM}}$ & $-0.7$ \\
    \bottomrule
    \end{tabular}
\end{table*}

\subsubsection{From SED fit to hyperspectral galaxy model}

\pkg{cigale} is run using PS1 filter transmission curves from
\cite[see Fig.~\ref{fig:cig_spectral_check}]{ps1photo} on photometric
pixels for which the Signal-to-Noise Ratio (SNR) is above 3 in all 5
bands.  Otherwise, the output flux is set to 0 at all wavelengths:
such pixels presumably belong to the sky or diffuse backgrounds, and
cannot be properly modeled by the SED fitter.

For all fitted pixels, \pkg{cigale} returns a spectrum
over an extended wavelength domain (from far UV to radio), with an
inhomogeneous spectral sampling between 1 and 5~\AA/px.  All spectra are
rebinned at the SEDm spectral sampling of $\sim 26$~\AA/px and truncated to the
$[3700, 9300]$~\AA{} range, resulting in 220~monochromatic slices.

\begin{figure}
  \centering
  \includegraphics[width=\linewidth]{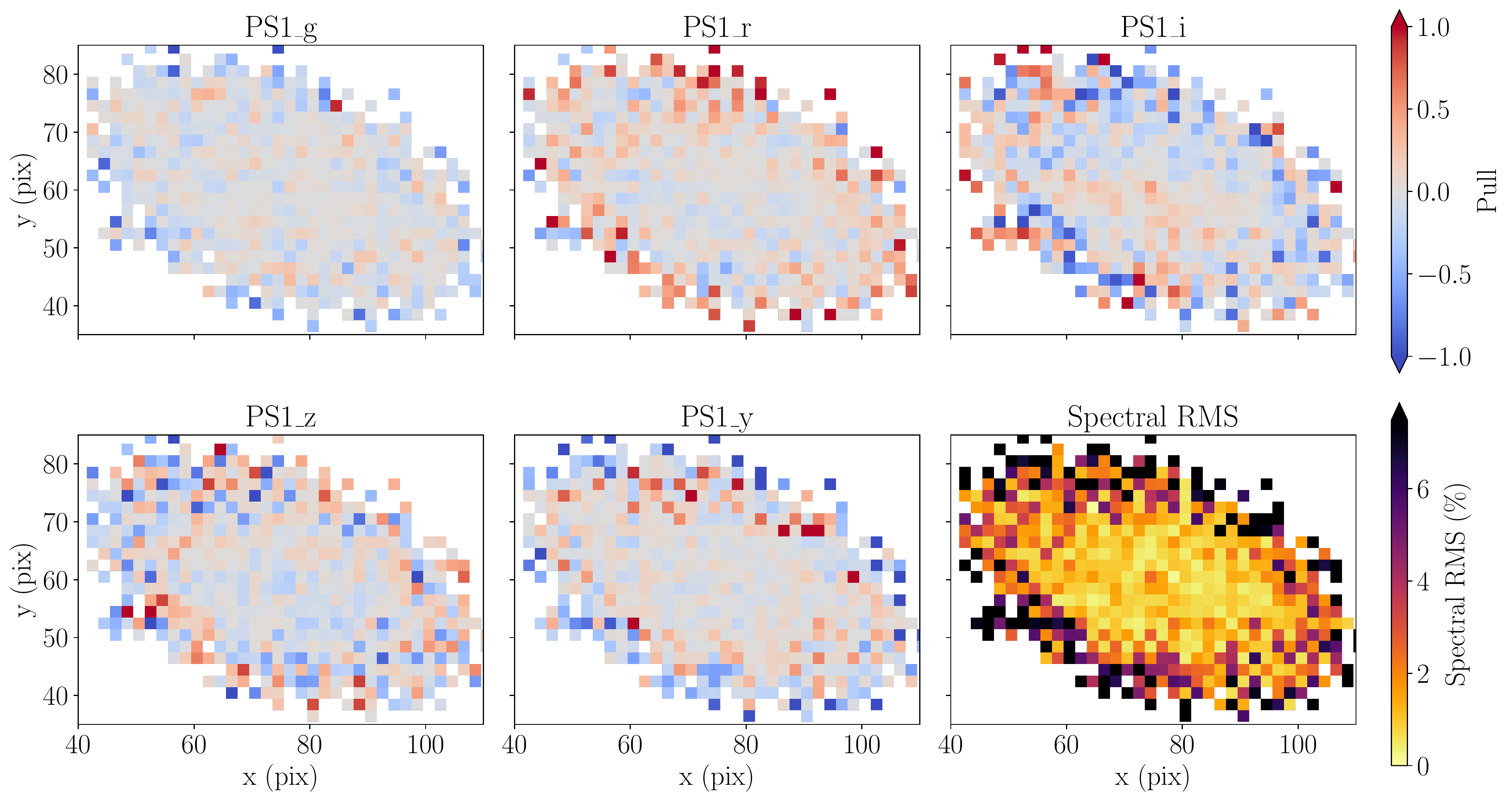}
  \caption{From \emph{left} to \emph{right} and \emph{top} to
    \emph{bottom:} map of the pull for the $grizy$ broadband images
    from \pkg{cigale} outputs, and spectral relative RMS over the 5
    reference host images.  Only pixels with SNR>3 for all $grizy$
    bands are considered (see Sec.~\ref{sec:sedfit}).}
  \label{fig:cig_spatial_check}
\end{figure}

The broadband flux from the SED fit is compared to the input
photometric measurements in Fig.~\ref{fig:cig_spatial_check}, where is
shown, for each PS1 band and pixel, the pull (i.e. the model residual
normalized by the error on the data) and the relative RMS averaged
over the 5 bands:
\begin{equation}
  \label{eq:RMS}
  \text{RMS} = \sqrt{\frac{1}{5}\sum_{\lambda=grizy} \left(
      \frac{f_{\lambda}-\tilde{f}_{\lambda}}{f_{\lambda}} \right)^2}
\end{equation}
where $f_{\lambda}$ denotes the data and $\tilde{f}_{\lambda}$ the predicted value.  The
averaged RMS is generally lower than 3\% in the core of the galaxy,
but can reach $\sim 10\%$ in the outer parts.  However, as the PS1
observations are 2 to 3 magnitude deeper than the SEDm ones
\citep{ChambersPanstarrs}, relatively poorly fitted pixels far away
from the host core have a marginal flux impact proportionately to the
SEDm background, and do not significantly affect the transient
spectrum in the scene model.

\subsection{SEDm Impulse Response Functions}
\label{ssec:irf}

The ``intrinsic'' hyperspectral galaxy model obtained from the SED fit
now has to be projected in the SEDm observation space, including the
spectro-spatial IRFs.  This section first presents the spectral
component, i.e. the Line Spread Function (LSF), then the spatial
component, aka the Point Spread Function (PSF).

\subsubsection{Spectral IRF (LSF)}
\label{sssec:lsf}

The output spectra from \pkg{cigale} have a spectral resolution of
$\sim 3$~\AA{} in the wavelength range 3200 to 9500~\AA{} (i.e. a
median resolving power of
$\mathcal{R} = \lambda/\Delta\lambda \sim 2000$, \citealt{bc03}),
$20\times$ the near constant SEDm resolution ($\mathcal{R} \sim 100$,
\citealt{SEDm}).  The full SEDm LSF is therefore a very good
approximation of the differential spectral IRF between \pkg{cigale}
and the SEDm.

To characterize the SEDm LSF, we use the intermediate line
fits of the wavelength solution derived from arc-lamp observations,
Cd, Hg, and Xe \citep[Sec.~2.1.2]{pySEDm}.  Each emission line is
fitted by a single Gaussian profile over a 3rd-order
polynomial continuum.

\begin{figure}
  \centering
  \includegraphics[width=\linewidth]{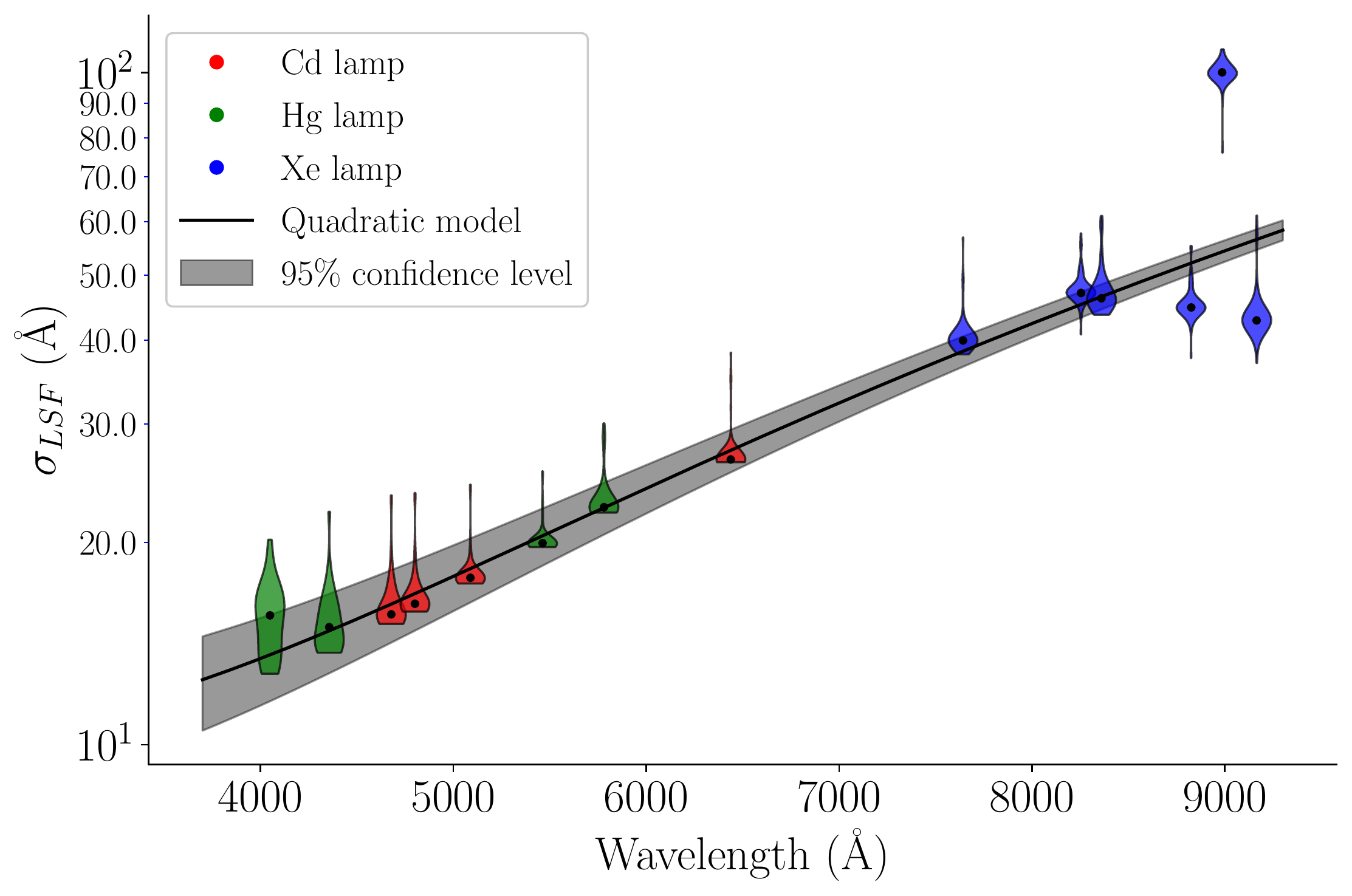}
  \caption{LSF standard deviation $\sigma_{LSF}$ as a function of
    wavelength, from the wavelength calibration of 65 nights between
    2018 and 2022.  Each violin corresponds to an emission line in the
    arc-lamp spectra (color legend).}
  \label{fig:lsf_sedm}
\end{figure}

Studying wavelength calibration for 65 nights between 2018 and 2022, the LSF
standard deviation $\sigma_{LSF}$ turned out to be stationary (no evidence of evolution
with time), fairly homogeneous in the FoV, but chromatic (as expected).
Figure~\ref{fig:lsf_sedm} shows the chromatic evolution of the standard
deviation, and the quadratic polynomial model adjusted to it.

\begin{figure*}
  \includegraphics[width=\textwidth]{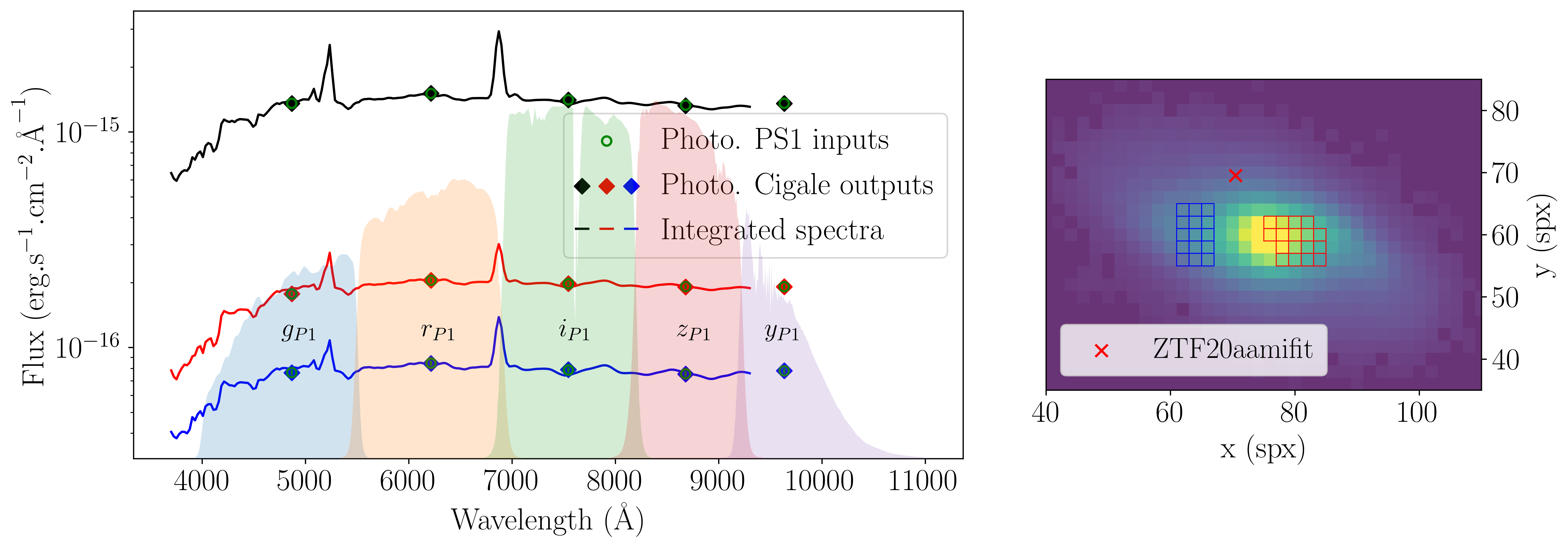}
  \caption{Hyperspectral galaxy model of ZTF20aamifit host galaxy,
    after projection in the SEDm observation space (including LSF).
    The \emph{green circles} correspond to the spatially integrated
    flux from PS1 cutouts, the \emph{black diamonds} to the same
    quantities as fitted by \pkg{cigale}.  The 5 shaded curves show
    the transmission of the $grizy$ PS1 filters.  The red and blue
    spectra on the left correspond to the spectra integrated in
    selected regions of same color in the model cube on the right; the
    black spectrum is the spectrum integrated over the full
    FoV.}
  \label{fig:cig_spectral_check}
\end{figure*}

To adapt the \pkg{cigale} spectra to the SEDm resolution, the spectra
of the hyperspectral galaxy model are convolved by the chromatic
Gaussian LSF.  An illustration of the result is shown in
Fig.~\ref{fig:cig_spectral_check}.

\subsubsection{Spatial IRF (PSF)}
\label{sssec:psf}

SNe are effective point sources, therefore solely described in the FoV
by the SEDm PSF (and its amplitude).  \hypergal uses a bisymetric PSF
model, in which radial profile is the sum of a Gaussian
$\mathcal{N}(r; \sigma)$ for the core, and a Moffat
$\mathcal{M}(r; \alpha, \beta)$ for the wings \citep{Buton2013,
  Rubin22}:
\begin{equation}
  \mathcal{P}(r; \alpha, \sigma, \beta, \eta) =
  \eta\times\mathcal{N}(r;\sigma) + \mathcal{M}(r;\alpha,\beta),
\end{equation}
where $r$ is an elliptical radius:
\begin{equation}
  \label{eq:ellipticity}
  r^{2} = (x-x_{0})^{2} + \mathcal{A}(y-y_{0})^{2} +
  2 \mathcal{B}(x-x_{0})\times(y-y_{0})
\end{equation}
with $(x_{0}, y_{0})$ the coordinates of the point source.  Parameters
$\mathcal{A}$ and $\mathcal{B}$ simultaneously describe the flattening
and the orientation of the PSF.

The 4 shape parameters $(\alpha, \beta, \sigma, \eta)$, which could be
ill-constrained in low SNR regime if adjusted independently, are
correlated by fixed relationships.  The PSF was tested on 148 isolated
standard stars, observed in 2021 with the SEDm, and we settled on the
following model.  The constrained PSF only has 2 free parameters:
$\alpha$ (Moffat radius) and $\eta$ (relative normalization of the
Gaussian), while the two other parameters are expressed as linear
functions of $\alpha$:
\begin{align}
  \beta &= \beta(\alpha) = \beta_0 + \beta_1 \alpha \\
  \sigma &= \sigma(\alpha) = \sigma_0 + \sigma_1 \alpha
\end{align}
where $\beta_0 = 1.53$, $\beta_1 = 0.22$, $\sigma_0 = 0.42$ and
$\sigma_1 = 0.39$ were determined from the training star sample.


The chromaticity of $\alpha(\lambda)$ is set as a power law function:
\begin{equation}
  \alpha(\lambda) = \alpha_{\sref}\left(\frac{\lambda}{\lambda_{\sref}}\right)^{\rho}
\end{equation}
where normalization $\alpha_{\sref}$ and index $\rho$ are free
parameters, and $\lambda_{\sref} \equiv 6000$~\AA.  Parameters $\eta$,
$\mathcal{A}$ and $\mathcal{B}$ do not exhibit strong chromaticity,
and are therefore considered constant.

Finally, the SEDm PSF of a given observation is fully described by 5
independent parameters: $\alpha_{\sref}$ and $\rho$, $\eta$,
$\mathcal{A}$ and $\mathcal{B}$.

\subsubsection{Differential PSF between PS1 and SEDm}
\label{sssec:relpsf}

The original hyperspectral galaxy model is derived from PS1
photometric exposures, with different seeing conditions than the SEDm
observations: the median seeing is $\sim 1\farcs7$ in SEDm
\citep{SEDm}, and $\sim 1\farcs2$ in PS1 images \citep{ps1pix}.

As the exact PSF profile is less critical for extended objects such as
the host galaxy, we chose to model the differential PSF between PS1
and SEDm as a single bisymmetric Gaussian kernel, with free
ellipticity and position angle.  The hyperspectral model is thus
convolved with this differential PSF before the spatial projection.

\subsection{Scene modeling}
\label{ssec:scene}


The two main elements are now at hand to build the scene model:
\begin{itemize}
\item a hyperspectral host galaxy model, and the (differential)
  spectral and spatial IRF to match it to the SEDm observations,
\item a chromatic PSF model for the transient point source.
\end{itemize}
The last component to complete the scene is the night sky and diffused light
background, modeled with a 2D 2nd-order polynomial at each wavelength.
The non-uniform terms handle a strong diffused light component, clearly visible
in the edges of the SEDm FoV and spectral range.  Overall, the background
component is described by 6~parameters, $b_{0}$, $b_{x}$, $b_{y}$,
  $b_{xy}$, $b_{xx}$ and $b_{yy}$.

We now describes the progressive method used to adjust it to the
observed SEDm cube, and the detailed spatial projection procedure used
to match the two cubes.

\subsubsection{General method}
\label{sssec:method}

We first consider $N \ll 220$ \emph{meta}slices of the SEDm cubes,
i.e. slices summed over a restricted wavelength domain, small enough
to be considered roughly achromatic, but large enough to increase the
SNR and significantly speed up the computation time.  The scene is
projected and fitted on all metaslices independently (the so-called
``2D fit'', Sec.~\ref{sec:fitter}), which results in a set of
$N \times m$ parameters; some are nuisance parameters (e.g. background
and component amplitudes), other key scene parameters, such as the
point source position and PSF shape parameters.

From this set of parameters evaluated at $N$ wavelengths, specific
chromatic models are used to fix all shape and position quantities
(the ``1D fit''), for which the full spectral resolution is not
required.  Ultimately, \hypergal performs a final linear ``3D'' fit of
the different component amplitudes over all monochromatic slices,
providing the total scene model cube at original SEDm spectral
sampling.

The pipeline uses by default $N=6$~metaslices linearly sampled between
5000 and 8500~\AA.  This spectral range is where the SEDm efficiency
is higher than 70\% \citep{SEDm}, and is extended enough to well
constrain the chromatic parameters, especially the ADR (see
Sec.~\ref{sec:chrom_fit}).  The pipeline was tested with different
number of metaslices, but no significant difference was noticed in the
results.

\hypergal was extensively optimized with the parallel computing
library \pkg{DASK}\footnote{\url{https://www.dask.org}} \citep{dask},
a dynamic task scheduler working as well on single desktop machines as
on many-node clusters. \pkg{DASK} optimizes the pipeline by analyzing
the (minimal) interdependencies between all computation tasks and
building an optimal parallelized workflow to be submitted and run on
an arbitrary number of available workers (in our case, we use 10 nodes
on the IN2P3 Computing Center\footnote{\url{https://cc.in2p3.fr/}}).

\subsubsection{Spatial projection}
\label{sec:projection}

The spatial projection of the hyperspectral galaxy model (matched to
the SEDm spectral and spatial IRFs) is made by successively projecting
each (meta)slice, taking into account the relative geometry and size
between PS1-derived model (square, $0\farcs 50$ aside) and SEDm
(hexagonal, $0\farcs 558$) spaxels.  The projection is done according
to a spatial anchor, a reference position in the sky supposedly known
in both (meta)slice.  The chosen anchor is the transient position,
derived from the ZTF survey astrometry and located at the center of
the queried PS1 images (and therefore at the center of the
hyperspectral model).  In the SEDm cube, this position is initially
guessed from the astrometric solution of the SEDm Rainbow Camera
\citep{SEDm, pySEDm}, but cannot be strictly fixed: the (chromatic)
SEDm anchor position $(x_{0}, y_{0})$ is free in the fitting process
of each metaslice.

The projection is made by geometrically overlapping the two polygonal
spaxel grids, with the anchor position as a reference; this is
effectively equivalent to a nearest neighbor interpolation scheme.
These computations are done using \pkg{shapely}%
\footnote{\url{https://github.com/Toblerity/Shapely}} \citep{shapely}
and \pkg{geopandas}%
\footnote{\url{https://github.com/geopandas/geopandas}}
\citep{geopandas}.

At this point, the model cube on which the PS1/SEDm differential PSF
and the SEDm LSF are applied is now projected in the SEDm observation
space, over the SEDm spaxel grid.

\subsubsection{Metaslice (2D) fit}
\label{sec:fitter}

As already mentioned, all components of the scene are first
independently fitted on the $N$ metaslices.  The free parameters per
metaslice are:
\begin{itemize}
\item the SN position $(x_{0}, y_{0})$ in the SEDm FoV, used as an
  anchor position for the spatial projection;
\item the SN PSF parameters ($\alpha$, $\eta$, $\mathcal{A}$,
  $\mathcal{B}$);
\item the PS1/SEDm differential PSF parameters ($\sigma_G$,
  $\mathcal{A}_G$, $\mathcal{B}_G$);
\item the amplitudes of the SN ($I$) and host ($G$) components;
\item the background coefficients
  ($b_{0}, b_{x}, b_{y}, b_{xy}, b_{xx}, b_{yy}$).
\end{itemize}
We use \pkg{iminuit}%
\footnote{\url{https://github.com/scikit-hep/iminuit}} \citep{minuit,
  iminuit2} to minimize a weighted $\chi^2$ for each metaslice
independently:
\begin{equation}
  \chi^2 = \sum\limits_{i}\left(\frac{y_i - \tilde{y}_i}{\sigma_{i}}\right)^2,
\end{equation}
where $i$ runs on the spaxels of the metaslice, $y$ and $\tilde{y}$
are the data and model fluxes respectively, and $\sigma$ the error on
the data.

\begin{figure*}
  \includegraphics[width=\textwidth]{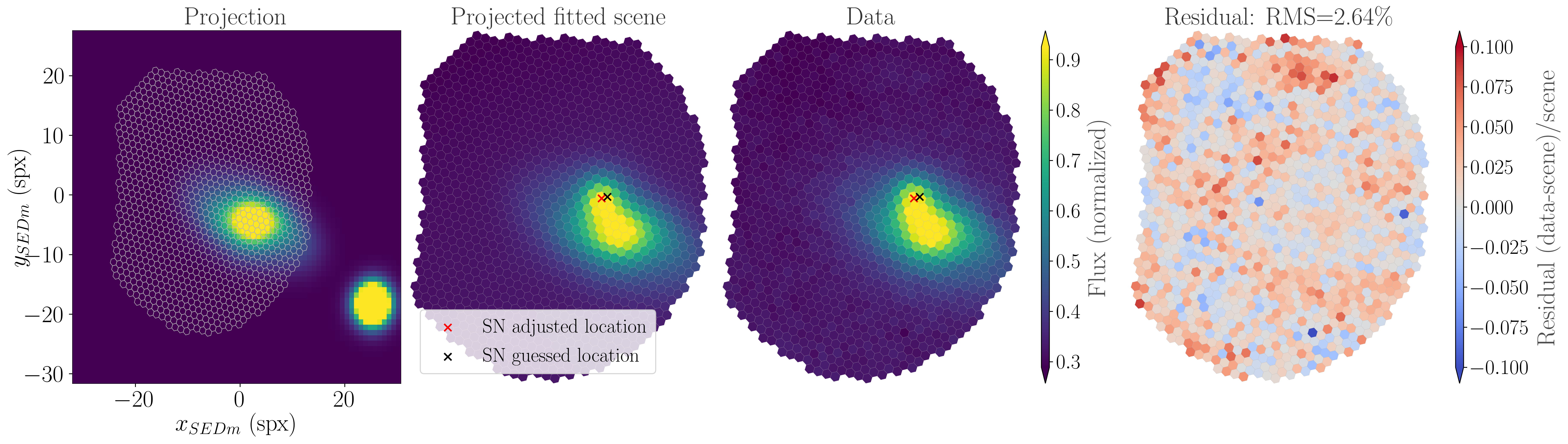}
  \caption{Fit result for the $[6167,6755]$~\AA{} metaslice of ZTF20aamifit
    cube.  \emph{From left to right:} metaslice from the original
    (transient-free) hyperspectral model with MLA footprint overplotted,
    projected fitted scene (host + background + SN), SEDm observations, and
    relative model residuals.}
  \label{fig:projection}
\end{figure*}

Fig.~\ref{fig:projection} illustrates the projection of one metaslice of the
hyperspectral galaxy model onto the SEDm space.  The fitted scene on this
metaslice shows a spatial RMS between the model and the data of 2.6\%.
Although indicative of the overall scene model accuracy, a low RMS does not
necessarily imply a clean separation of the different components, (e.g. when
the transient lies on top of a sharp host galaxy core).  Extraction accuracy is
directly evaluated from simulated SN spectra in Sec.~\ref{sec:validation}.

\subsubsection{Chromatic (1D) fit}
\label{sec:chrom_fit}

Once the fit is performed independently over all $N$ metaslices, a set
of $N$ chromatic estimates of the $m$ parameters is at hand, and used
to assess their (smooth) chromatic evolution (except for the component
amplitudes and background parameters, which are nuisance parameters at
this point).

The chromaticity of the full Gaussian + Moffat PSF is modeled as
detailed in Sec.~\ref{sssec:psf}.  The chromaticity of the width of
the 2D Gaussian which models the differential PSF between PS1 and the
SEDm is adjusted by a similar power law,
\begin{equation}
  \sigma_G(\lambda) =
  \sigma_{\sref} \left(\frac{\lambda}{\lambda_{\sref}}\right)^{\rho_G}
\end{equation}
where $\rho_G$ and $\sigma_{\sref}$ are adjusted on the $N$ metaslice
estimates obtained previously, and $\lambda_{\sref} \equiv 6000$~\AA;
the shape parameters $\mathcal{A}_G$ and $\mathcal{B}_G$ are
considered constant equal to their (inverse-variance weighted) mean
values over the $N$ metaslices.

The effective anchor location in the SEDm FoV is systematically
wavelength-dependent, due to the chromatic light refraction through
the atmosphere (ADR).  Given the $N$ positions of the SN in the
different metaslices, an effective 4-parameter ADR can be
fitted to
track the chromatic offsets in the FoV:
\begin{equation}
  \label{eq:adr}
  \begin{bmatrix}
    x_0(\lambda) \\
    y_0(\lambda)
  \end{bmatrix} =
  \begin{bmatrix}
    x_{\sref} \\
    y_{\sref}
  \end{bmatrix} - \frac{1}{2}\left(
    \frac{1}{n^{2}(\lambda)} - \frac{1}{n^{2}(\lambda_{\sref})}\right)
  \times \tan(d_{z})
  \begin{bmatrix}
    \sin\theta \\
    \cos\theta
  \end{bmatrix}
\end{equation}
with ($\theta$, $z$, $x_{ref}$, $y_{ref}$)
  the fitted parameters, where $\theta$ is the parallactic angle, $z$ the airmass and
$d_z=\arccos{z^{-1}}$ the zenith distance in the plane-parallel
atmosphere approximation, and $(x_{\sref}, y_{\sref})$ the reference
position at reference wavelength $\lambda_{\sref} \equiv
6000$~\AA.
The index of refraction $n(\lambda)$ of air is computed using the
Edlén equation from \citet{refractindex}%
\footnote{\url{https://emtoolbox.nist.gov/Wavelength/Documentation.asp}},
which takes into account the atmospheric pressure, temperature, and
relative humidity, as provided for each exposure by the SEDm Telescope
Control System.

\begin{figure}
  \centering
  \includegraphics[width=\linewidth]{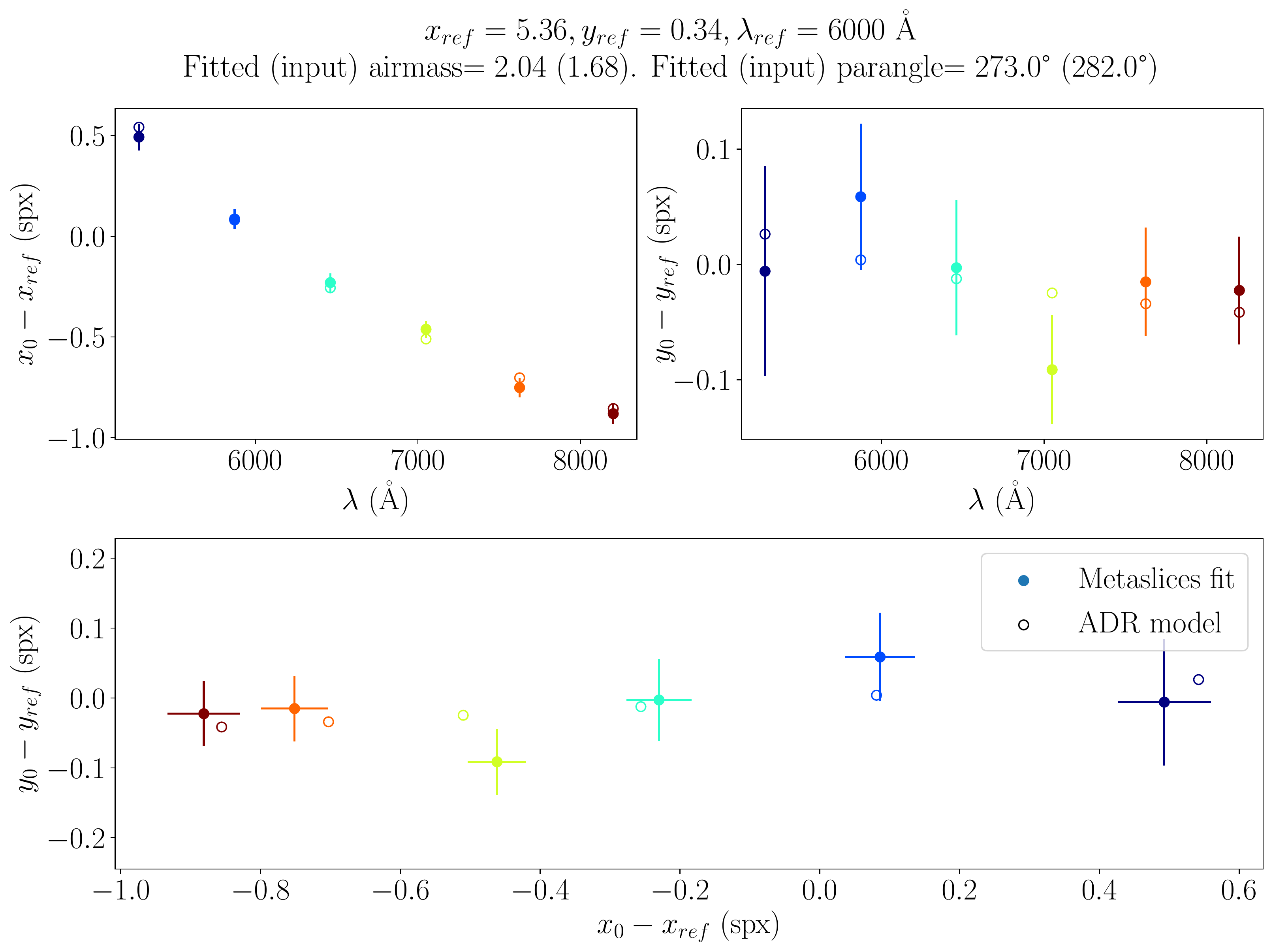}
  \caption{SN positions as a function of wavelength, and the effective
    ADR fit. \emph{Top panel}: relative offsets with respect to
    reference position at reference wavelength along each axis; filled
    points correspond to the observed offsets, and open circles to the
    predictions of the ADR model.  \emph{Bottom panel}: relative
    offsets in the $(x,y)$ plane.  Color codes for the central
    wavelength of the metaslices.}
  \label{fig:output_adr}
\end{figure}

Figure~\ref{fig:output_adr} illustrates the ADR effect, a drift of the
metaslice anchor position with wavelength, and the ADR model, at
effective airmass $\sim 2.0$.

\subsubsection{Final (3D) fit}
\label{sec:3Dfit}

Once all PSF and ADR chromatic models are available from 2D+1D metaslice
adjustments, the scene morphological parameters are considered known
  and fixed at each wavelength: the point source position $(x_{0}, y_{0})$ and
  PSF parameters ($\alpha$, $\eta$, $\mathcal{A}$, $\mathcal{B}$), as well as
  the PS1/SEDm differential PSF parameters ($\sigma_G$, $\mathcal{A}_G$,
  $\mathcal{B}_G$).  This allows us to perform a final 3D linear fit over all
monochromatic slices, where only scaling amplitudes of the different scene
components -- namely host galaxy $\{G\}$, SN $\{I\}$ and background
  polynomial components $\{b_{0}, b_{x}, b_{y}, b_{xy}, b_{xx}, b_{yy}\}$ --
are let free per slice. The total scene is then reconstructed at full
spectral resolution.

Although $G(\lambda)$ is primarily used to recover flux calibration
mismatch between PS1 and SEDm, this normalization parameter can interfere
in a non-trivial way with the position and intensity of the emission
lines in the hyperspectral galaxy model. This effect might help to handle
slightly incorrect input redshift used in the SED fitting step, especially under the assumption of a
uniform spatial distribution of the line. As this has not been analysed
in depth, we extend this thought in Sec.\ref{sec:discuss}.

\begin{figure}
  \centering
  \includegraphics[width=\linewidth]{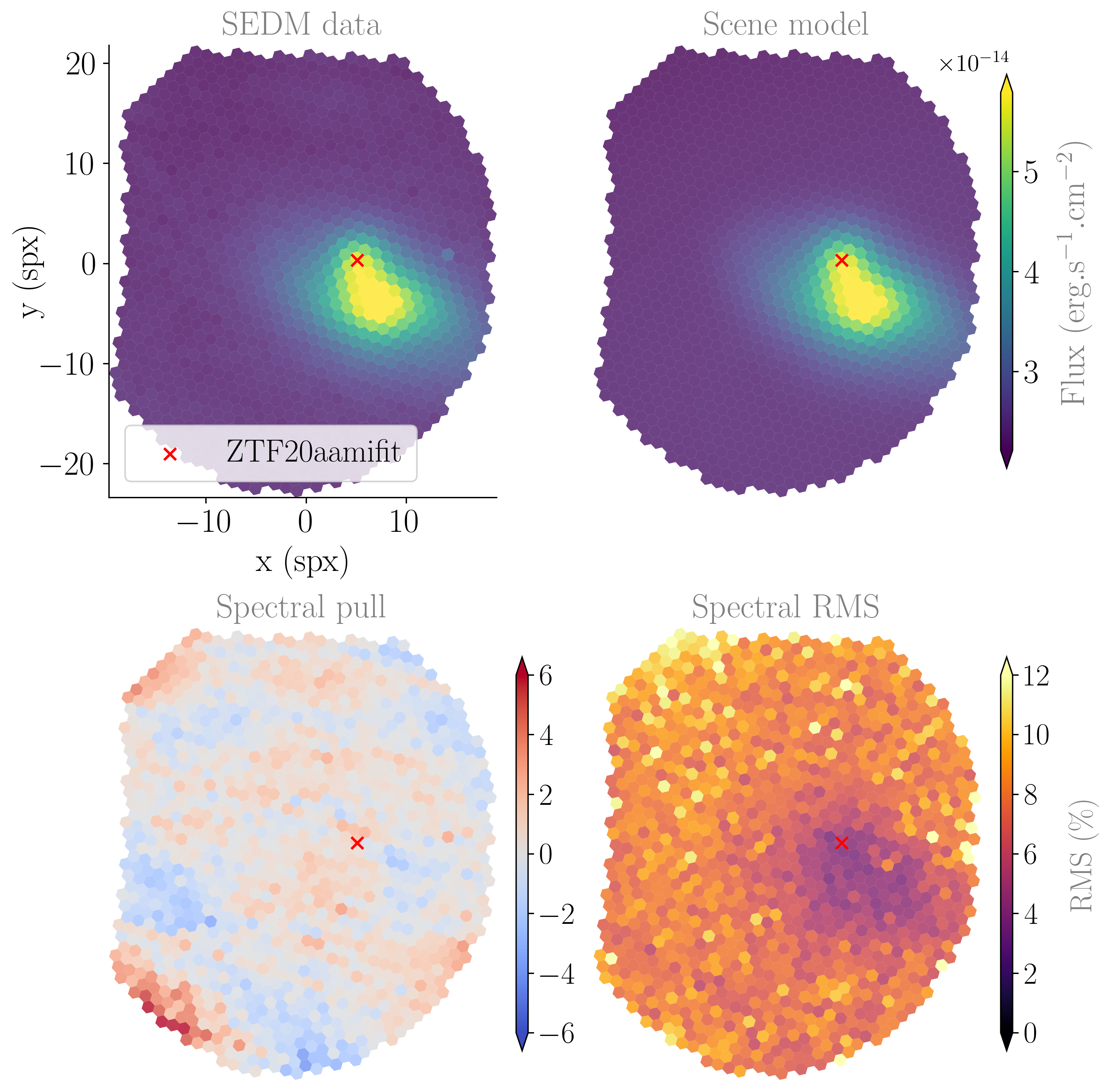}
  \caption{Full scene model for ZTF20aamifit. \emph{Top panel}:
    integrated SEDm and \hypergal-modeled cubes; the red cross
    indicates the adjusted point source position at 6000~\AA.
    \emph{Bottom panel}: spectral pull and spectral relative RMS.  No
    galaxy- or SN-related structured residual is visible in the pull
    map and the spectral RMS indicates an accuracy of $\sim 4\%$ at
    the host and SN locations.}
  \label{fig:output_global}
\end{figure}

Fig.~\ref{fig:output_global} presents the white image (spectral
integral) of the final \hypergal scene model for SN ZTF20aamifit.  The
quality of the fit is evaluated from the pull map, showing no evidence
of structured residuals.  The spectral relative RMS map indicates an
accuracy of $\sim 4\%$ at SN and host core location, and 6 to 7\%
where only the background is significant.

\subsection{Component extraction}
\label{sec:source_extract}

\begin{figure}
  \centering
  \includegraphics[width=\linewidth]{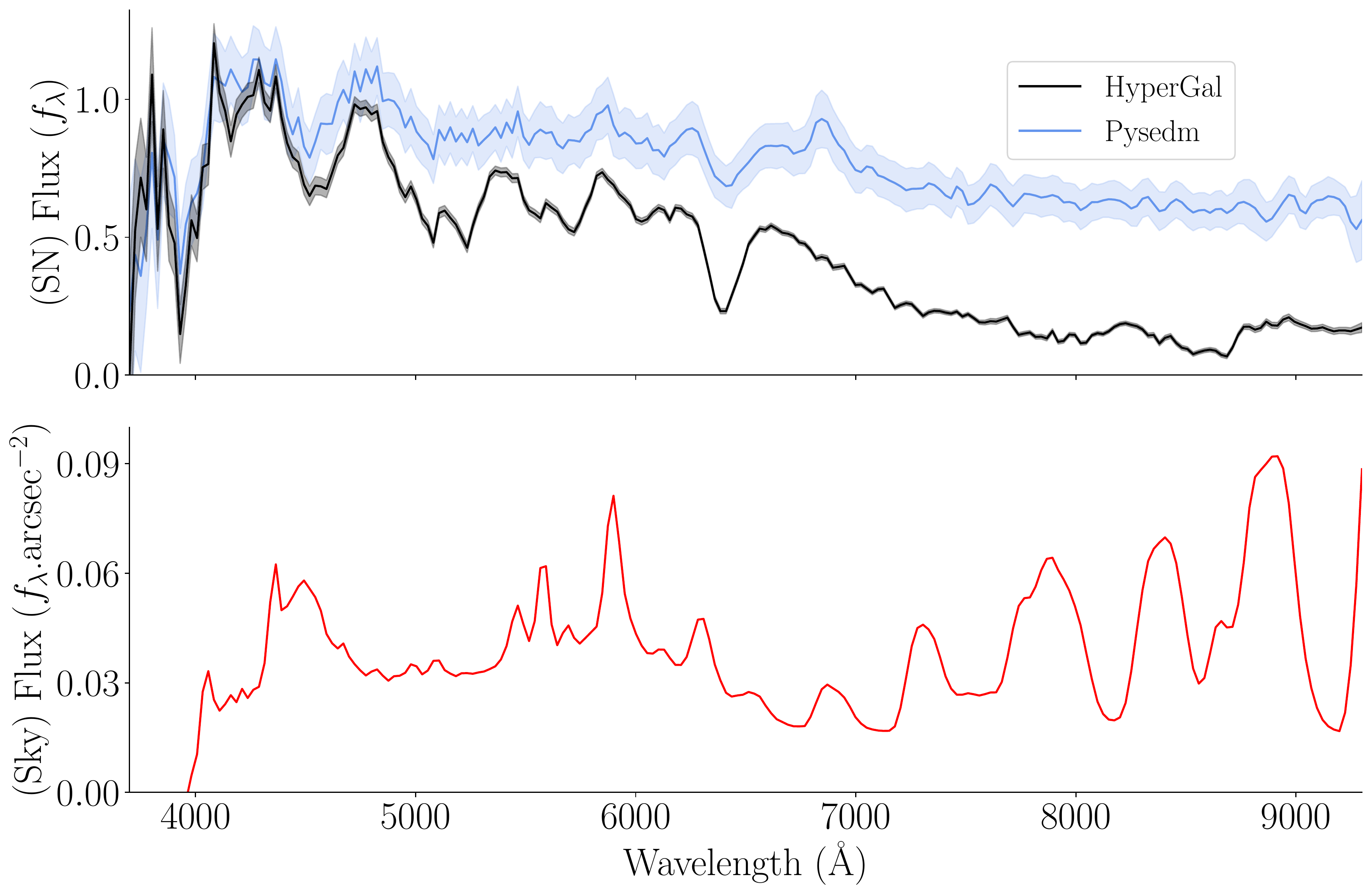}
  \caption{SN ZTF20aamifit spectrum -- as extracted by \hypergal (\emph{black})
    and \pkg{pysedm} (\emph{blue}) -- and uniform sky spectrum (coefficient
    $b_0(\lambda)$, \emph{red}).  Flux unit $f_{\lambda}$ stands for
    femto-\si{erg.cm^{-2}.s^{-1}.\AA^{-1}}.}
  \label{fig:outputsnspec}
\end{figure}

The strength of the \hypergal pipeline is the simultaneous fit of the
3 scene components, the host galaxy, the transient point source and
the background.  The main quantity of interest is of course the SN
spectrum (i.e. the vector of the point source amplitudes $I(\lambda)$,
see Fig.~\ref{fig:outputsnspec}), but one can also selectively
subtract individual components to assess the quality of the scene
model.

\subsubsection{Host galaxy integrated spectrum}
\label{sec:hg_outputs}

\begin{figure*}
  \includegraphics[width=\textwidth]{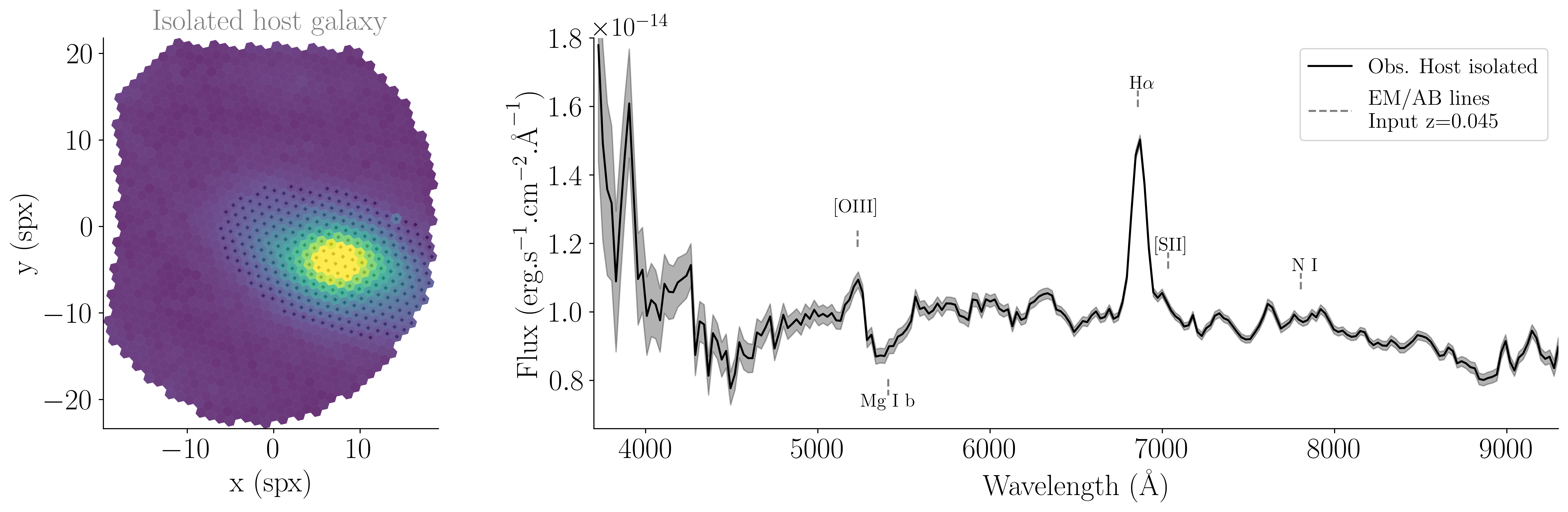}
  \caption{ZTF20aamifit host galaxy, isolated from the SEDm data cube.
    \emph{Left panel:} isolated host galaxy component in the SEDm
    cube, after subtraction of both the SN and background models.
    \emph{Right panel:} host spectrum integrated over the selected
    spaxels; the main spectral features are marked for the input
    redshift $z=0.045$.}
  \label{fig:output_host}
\end{figure*}

Thus, the host contribution can be isolated in the SEDm cube by
subtracting the SN and the background components (see
Fig.~\ref{fig:output_host}).  To further compute an integrated host
spectrum, a large elliptical aperture is defined around the host with
the \pkg{SEP} package \citep{sep, sep2} from the PS1 images.
This aperture is then projected in the SEDm cube, using the respective
World Coordinate Systems.  Note that the ADR is neglected in the
process, as it rarely induce a deviation of more than one or two
spaxels in the FoV, and has barely any impact on the host spectrum
integrated over a large aperture.

The integrated host spectrum is shown in Fig.~\ref{fig:output_host},
with the expected position of some major emission lines at the input
redshift (independently of the host spectrum).  This procedure
highlights the consistency between the input redshift used for the
hyperspectral galaxy modeling and the extracted integrated spectrum.
In the future, it could be considered to consistently estimate the
host redshift directly from such integrated spectrum during the scene
modeling (see Sec.~\ref{sec:discuss}).

\subsubsection{Point source radial profile}

\begin{figure}
  \centering
  \includegraphics[width=\linewidth]{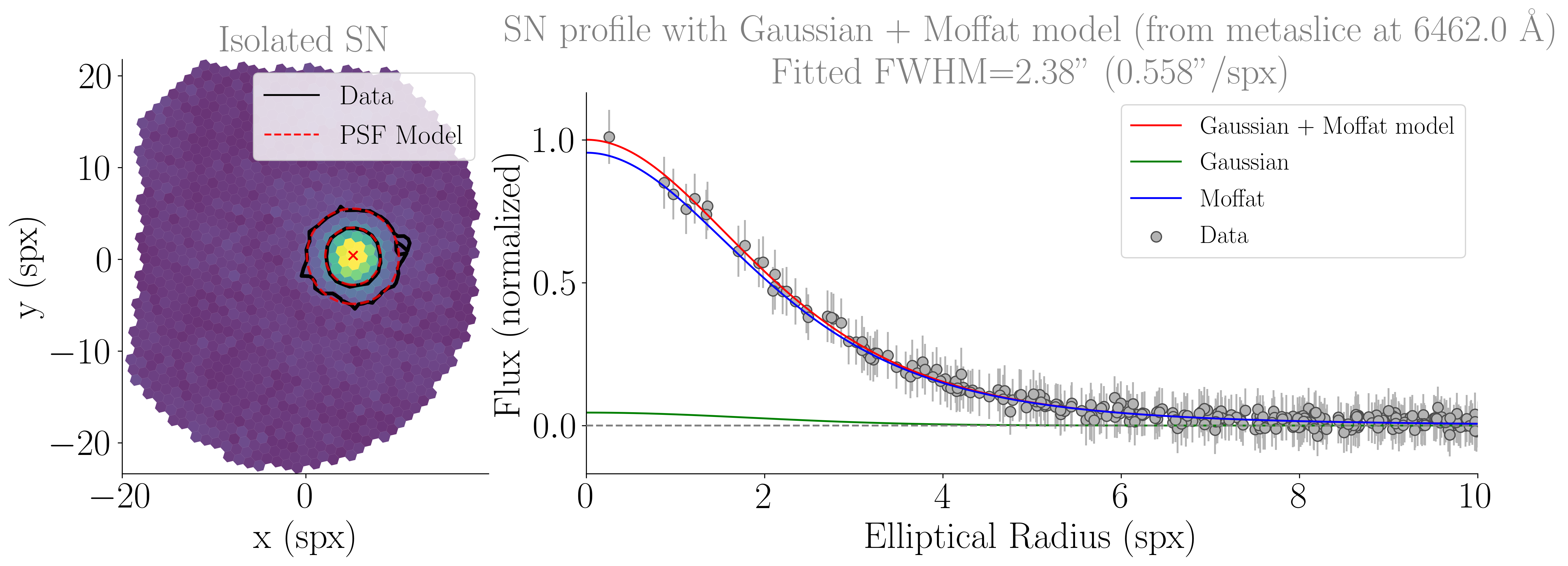}
  \caption{SN ZTF20aamifit, isolated from the SEDm data cube.
    \emph{Left panel}: isolated SN component in the SEDm
    $[6167,6755]$~\AA{} metaslice, after subtraction of both the host
    and background models; the \emph{red cross} indicates the
    fitted
    SN location, and \emph{contours} show the elliptical iso-radius at
    3 and 5~spx for observations (\emph{black solid lines}) and model
    (\emph{red dashed lines}). \emph{Right panel}: PSF profile for the same
    metaslice, as a function of the elliptical radius.  The data
    points refer to the isolated SN on the left panel, the \emph{red
      curve} corresponds to the PSF profile (without the background),
    the \emph{blue} and the \emph{green curves} to the Moffat and the
    Gaussian components respectively.  The Gaussian component is
    particularly weak because of the poor seeing conditions.}
  \label{fig:output_sn_ifu}
\end{figure}

Similarly, the point source contribution can be isolated in the SEDm
cube by subtracting both host and background models, as shown in
Fig.~\ref{fig:output_sn_ifu} for the $[6167,6755]$~\AA{} metaslice of
the ZTF20aamifit cube.  This closer look at the point source
contribution allows us to check the accuracy of the PSF profile in
each metaslice.  The fact that the profile smoothly tends to 0 means
that the background was correctly modeled by \hypergal; also, the
absence of outliers in the data points indicates that there is no
evidence of residual host contamination in the profile, as noticed in
the isolated SN image.

\subsection{SN classification}

\hypergal being primarily designed for the transient spectral
classification, an automated typing procedure is included in the
pipeline, based on the Supernova Identification
\cite[\pkg{SNID}][]{snid}.  The typing is performed over the 4000 to
8000~\AA{} spectral range, which includes the most discriminating
spectral features for redshifts $z \lesssim 0.1$.  This domain also
corresponds to the one where the SEDm CCD quantum efficiency is
over~60\%.

The quality of the \pkg{SNID} classification is quantified by the
r$lap$ parameter, measuring the strength of the correlation between
the input and template spectra.  According to \citet{snid}, an
r$lap \geq 5$ indicates a high confidence in the classification,
without considering any prior on the redshift or the phase of the SN.

\begin{figure*}
  \includegraphics[width=\textwidth]{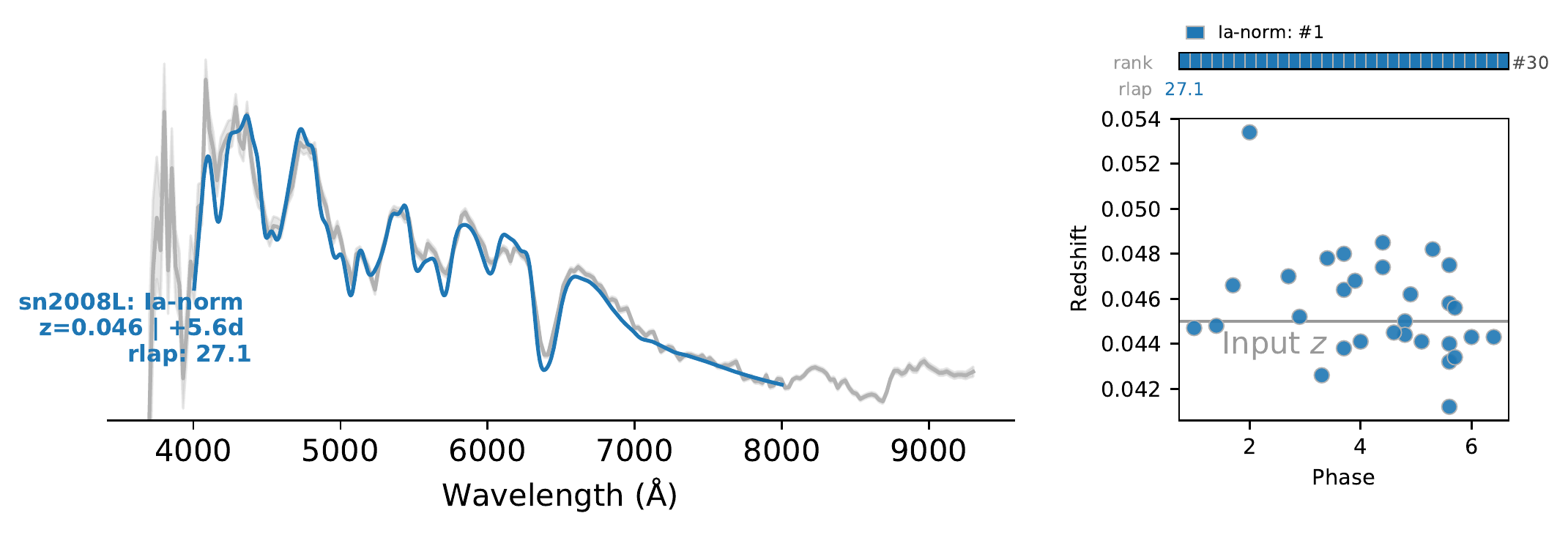}
  \caption{\pkg{SNID} typing of the ZTF20aamifit \hypergal spectrum.
    \emph{Left panel}: input spectrum (in grey) and best model from
    \pkg{SNID} (in blue).  \emph{Right panel}: distribution in the
    (redshift, phase) plane of the 30 best matches with an r$lap > 5$
    (all being normal SNe~Ia in this case).  The input redshift of the
    galaxy ($z=0.045$) is indicated with the horizontal grey line.
    The best model, with a very high r$lap = 27$, classifies
    ZTF20aamifit as an SN~Ia at redshift ${z=0.046}$ and phase
    $p=+5.6$~days.}
  \label{fig:output_snid_typing}
\end{figure*}

Figure~\ref{fig:output_snid_typing} presents the \pkg{SNID} typing of
ZTF20aamifit using its \hypergal-extracted spectrum.  The best match
has an r$lap = 27$, which leaves no doubt about its classification as
an SN~Ia.  In comparison, the \pkg{pysedm}-extracted spectrum (see
Fig.~\ref{fig:outputsnspec}) is also typed as an SN Ia but with a
significantly lower confidence (r$lap = 9$).

\section{\hypergal validation}
\label{sec:validation}

The \hypergal pipeline is validated with a set of simulations, in order to
quantify the accuracy of the extracted SN spectra as a function of various
observational conditions, and the ability to spectrally classify the transient.
In this section, we first present the simulation process, before performing
some statistical analysis on the spectral accuracy, followed by the typing
efficiency.  For comparison, the SNe are also extracted with a method similar
to \pkg{pysedm} \citep{pySEDm}, that is a plain PSF extraction of a
  supposedly isolated source (not accounting for the background galaxy), but
  using the same PSF and diffuse background models as \hypergal for
  consistency.

\subsection{Simulated sample}

During a short shutdown of the main ZTF camera, the SEDm was free to observe a
few galaxies which hosted SNe at least one year earlier.  These observed host
cubes are therefore naturally in the SEDm space for which \hypergal is
designed; 10 different hosts with various morphologies were acquired at
different locations in the IFU and with an airmass ranging from $1.01$ to
$2.04$.  This allows us to cover a large variety of observation conditions,
from the ideal case to the poorest condition.  An
artificial point source, whose spectrum and type is known a priori, is then
added to these cubes.

To mimick SEDm spectra as much as possible, we use spectra of
well-isolated transients observed with SEDm and successfully
classified by \pkg{SNID} with a very high r$lap$.  For the SNe~Ia (the
most numerous to be observed), 70 spectra are selected with
r$lap > 25$ for the best model and r$lap > 15$ for the first
30 models.  Similarly, 7 SNe~II spectra with r$lap > 12$ are selected.
For the more rarely observed SNe~Ic and SNe~Ib ($\sim 5\%$ of
observations), only one spectrum of each was chosen, but with high
classification confidence (r$lap \sim 22$ for the~Ib and
r$lap \sim 13$ for the~Ic).  To increase the SNR, each of these
spectra is then slightly smoothed using a Savitzky-Golay filter
(3rd-order polynomial over a window of 5 pixels), in order to keep
intact the spectral structures.

While building the simulated sample, the different SN types are distributed
to follow the observed fractions \citep{ztfspecred}, with 80\% of
SNe~Ia, 15\% of SNe~II, 2.5\% of SNe~Ib, and 2.5\% of SNe~Ic.  For further
analysis, Ib and Ic will be studied jointly as SNe~Ibc.

A marginalization on the phase of the SNe~Ia is applied, based on the
  DR1 statistics from the ZTF SN~Ia group \citep{Dhawan2022}.  Knowing the
  phase of the 70 SN~Ia input spectra used for the simulation, we draw the SN
  templates to follow the observed distribution of phases, modeled as a
Gaussian distribution centered on $-3$~days with a standard deviation of
4~days.

Concerning the PSF, the profile is assumed to follow the model
presented in the Sec.~\ref{sssec:psf}.  To faithfully represent the
seeing diversity of the observations, the chromatic radial profile
parameters are drawn from the joint distribution built from
$\sim 2000$~standard stars, thus taking into account the latent
correlations between parameters.

Finally, 2 extra parameters -- which we consider the most likely to impact on
the \hypergal robustness -- are introduced in the simulations: the
contrast $c$ between the transient and the local background, and the
distance $d$ between the target and the host.

The latter aims to cover all observed cases, from the exact overlapping between
the point source and the host ($d \approx 0$) to the limit of an unstructured
background ($d \gg$ host core size).  The host center is identified by matching
the WCS solution from the SEDm cube and the underlying photometric images from
PS1.  The distance $d$ is drawn from a uniform distribution between 0 and
$5\farcs6 \equiv 10$~spx.  As the SEDm mostly observes well centered point
sources, the simulated SN is placed within 12~spx from the center of the FoV,
or at least towards the MLA center if the host is on the edge.

The contrast $c$ is defined by $c = S/(S+B) \in[0,1]$ where $S$ is the
transient signal, and $B$ is the total (sky and host) background, both
spectrally integrated over the equivalent $r$ band of ZTF.  For a
random $c$ drawn from a uniform distribution in $[0, 1]$, the
background signal $B$ is first estimated at the simulated SN location,
by successively integrating spatially the pure host cube weighted by
the chromatic PSF profile, then spectrally over the ZTF $r$ band.
Once $B$ is known, the SN spectrum is scaled so that the $r$-band
integral $S = cB / (1 -c)$.  Finally, the simulated SN contribution to
the cube variance is added to the one from the host galaxy, under the
hypothesis of pure photon noise, using the flux solution of the host
cube.

Ultimately, the 5000~simulated cubes are built, covering a large range
  of observation conditions, host galaxy morphologies and positions in the FoV,
  transient locations and spectral types and SNR. The \hypergal pipeline and
  the standard point source extraction are then used to estimate the resulting
  SN spectra.

\subsection{Extraction accuracy}
\label{ssec:rmscontrast}

The SEDm is designed and used for the spectral classification of
transient.  Thus, beyond pure absolute spectro-photometric flux
accuracy, what is important is the capacity of \hypergal to extract
the spectral features allowing a proper classification, independently
of the absolute flux level or even the large-scale continuum shape.
Consequently, the \hypergal performances are evaluated on
continuum-normalized transient spectra in the $[4000,8000]$~\AA\
wavelength range, as in \pkg{SNID}.

The continuum is fitted as a 5th-order polynomial over the
wavelength range slightly extended by 100~\AA{} at each extreme, to
avoid some unwanted boundary effects.  The spectral comparison between
simulation input and \hypergal/standard method output spectra is then
systematically performed on continuum-normalized spectra, and is
quantified using a wavelength-averaged relative RMS similar to
Eq.~(\ref{eq:RMS}):
\begin{equation}
  \label{eq:RMS2}
  \text{RMS} = \sqrt{\frac{1}{N}\sum_{\lambda} \left(
      \frac{f_{\lambda}-\tilde{f}_{\lambda}}{f_{\lambda}} \right)^2}
\end{equation}
where $N$ refers to the number of monochromatic slices between
  $[4000,8000]$~\AA, $f_{\lambda}$ denotes the data and $\tilde{f}_{\lambda}$
  the predicted value.

\begin{figure*}
  \centering
  \includegraphics[width=\textwidth]{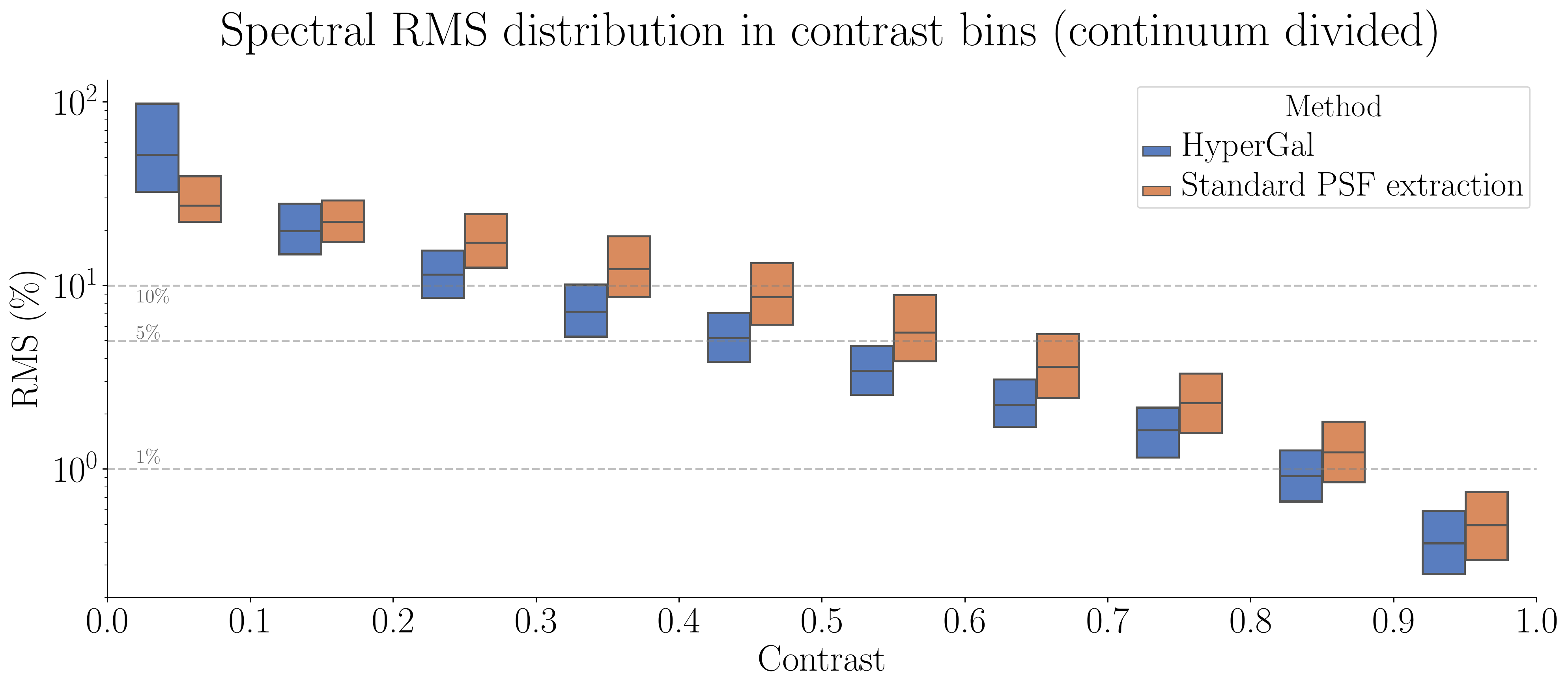}
  \caption{Distribution, as a function of the contrast, of the
    spectral relative RMS between simulation input spectra and
    extracted spectra, averaged over the $[4000, 8000]$~\AA{} domain.
    In the boxes, the 3 levels represent the 3 quartiles (25\%,
    median, and 75\%).  Each bin includes the same number of
    simulations, as the contrast $c$ is uniformly distributed in
    $[0, 1]$.}
  \label{fig:rmscontinuumdivide}
\end{figure*}

The distance $d$ is found to have no influence on the spectral
accuracy of \hypergal, with an absolute correlation coefficient lower
than 0.2.  On the other hand, Fig.~\ref{fig:rmscontinuumdivide} shows
the correlation between spectral relative RMS and contrast $c$ for
both extraction methods on continuum-normalized spectra.  The results
are marginalized over all SN types, as the extraction accuracy is
supposedly independent of the spectral shape.

Both methods obtain an RMS greater than 20\% for $c<0.2$, suggesting
that spectral classification at such low contrast will be difficult.
Yet, the standard method seems to be more accurate than \hypergal at
extremely low contrast ($c<0.1$); this actually appear to be an
artifact of the continuum normalization.  At very low contrast,
neither methods can reasonably disentangle the SN from the background;
however, by effectively mixing SN and host signal, the standard point
source extracted spectrum has a higher SNR (although less accurate),
and the continuum normalization is less prone to fail
catastrophically, contrary to the case of the spectrum consistent with
0 as extracted by \hypergal.

\hypergal starts to stand out for $0.2 < c < 0.3$ with a median RMS
around 10\%, and RMS decreases steadily below 10\% at $c > 0.3$, 5\%
for $c > 0.5$, and 1\% for $c > 0.8$.  Compared to the standard
extraction method, \hypergal shows a median improvement of $\sim 50\%$
for $0.2 < c < 0.6$, and gradually returns to a median improvement of
$\sim 20\%$ up to highest contrasts.  Since the continuum
normalization removes the effects of absolute scaling and color terms
on the spectral RMS, the improvement exclusively relates to the
contamination of the SN spectrum by the host galaxy spectral features.
This demonstrates the effectiveness of \hypergal to drastically reduce
this host contamination.

\subsection{Distribution of contrast in the observations}

Before turning to the classification efficiency, the contrast
distribution in the SEDm observations is estimated, as a reference to
compare our results with.  Rather than using \hypergal on observations
made with the SEDm (as was actually done for the ZTF Cosmology SN~Ia
Data Release 2 to come \citealt{dr2rigault}) -- this would be like
evaluating the pipeline with itself -- the contrast $c = S/(S+B)$ is
estimated from photometric images of the same DR2 sample, made up of
about 3000~SNe~Ia.

For each SN, its signal $S$ in the PS1 $r$ band at the date of the
SEDm observation is estimated from the SALT2 fit \citep{Guysalt2005,
  Guysalt22007, Betoule2014} of its light curve.  We chose the PS1 $r$
band, in practice very similar to the ZTF one, because only images
from this survey were available at the time of the study.  On the
other hand, the host contribution to the background $B_{gal}$ is
estimated from the integrated flux within a radius of $2\arcsec$
around the SN.  As PS1 images are already sky-subtracted, an
additional sky background $B_{sky}$ has to be added for a fair
comparison with simulations.  Two different values are used: a
fiducial value $m_{sky} = 20$~mag, approximatively corresponding to
the magnitude depth of the SEDm, and a more conservative value
$m_{sky} = 21$~mag.  The sky background being largely negligible in
front of a galactic one, its exact value essentially alters the high
contrast values: for a SN isolated from its host galaxy, the contrast
would systematically increases as the sky background tends to~0.

\begin{figure}
  \centering
  \includegraphics[width=\linewidth]{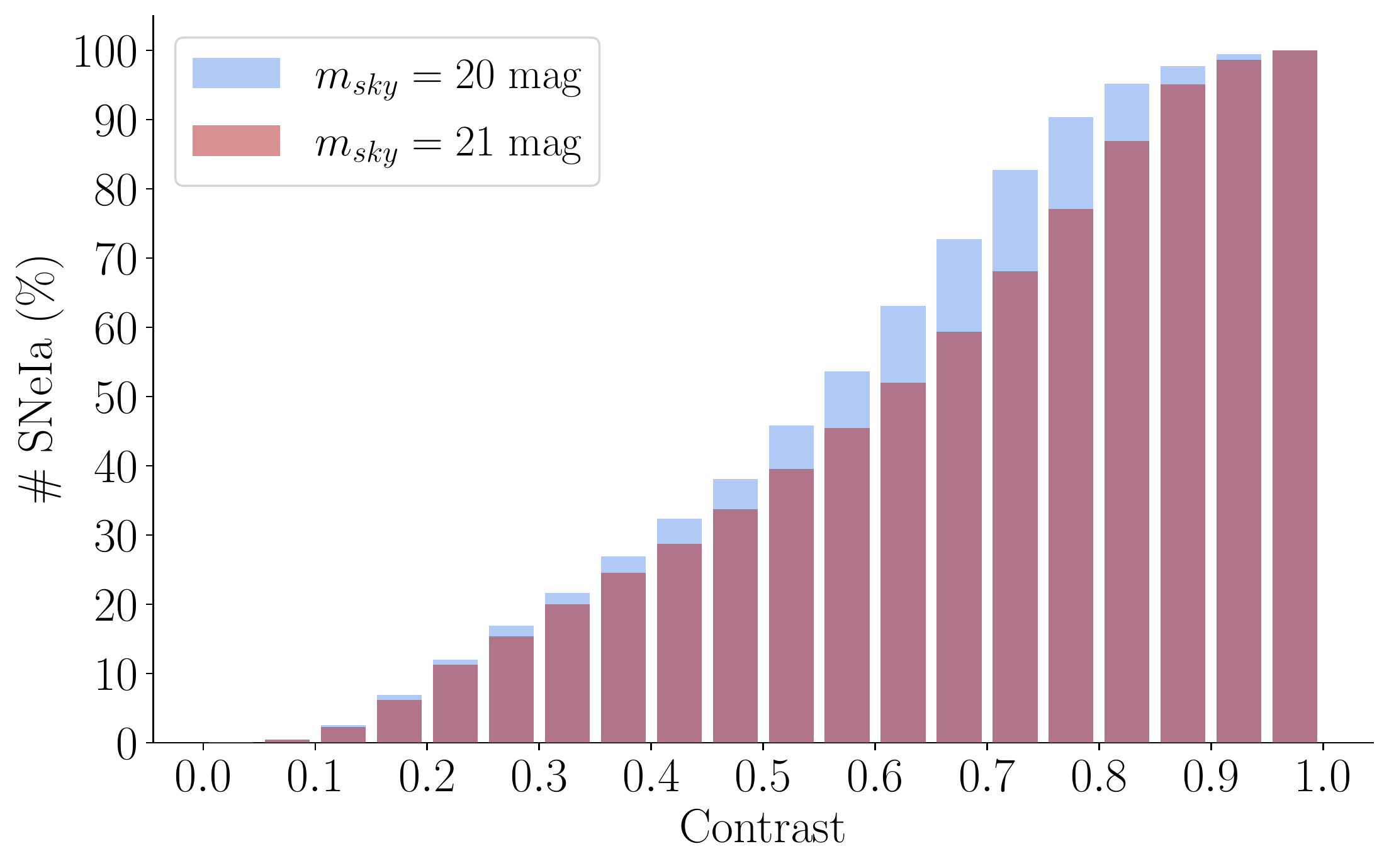}
  \caption{Cumulative contrast distribution estimated from $\sim 3000$
    SN~Ia observed with the SEDm.  Since only $B_{gal}$ is estimated
    from PS1 images, an additional $B_{sky}$ is estimated using two
    different sky levels, $m_{sky} = 20$ (\emph{blue}) for a realistic
    value, and $m_{sky} = 21$ (\emph{red}) for a conservative value.}
  \label{fig:contrastdist}
\end{figure}

Figure~\ref{fig:contrastdist} displays the cumulative distribution of
the contrast for the DR2.  The median contrast of this distribution is
$c=0.58$ for $m_{sky} = 20$~mag and $c=0.63$ for $m_{sky} = 21$~mag.
For both sky levels, less than 1\% of observations have a contrast
$c < 0.1$, and only 7\% with $c < 0.2$.  At high contrast end, 2 to
5\% of the observations have a $c > 0.9$ depending on the adopted sky
magnitude.  Almost 95\% of observations have a contrast
$0.1 \leq c \leq 0.9$, and a slightly less than 90\% with
$0.2 \leq c \leq 0.9$.

According to the results of Sec.~\ref{ssec:rmscontrast}, one can therefore
assess the spectral accuracy of \hypergal on the DR2 sample -- using
  the spectral relative RMS (Eq.~\ref{eq:RMS2}) as an indicator -- to be of
the order of 10\%, 5\% and 2\% for 80\%, 60\% and 20\% of the observations
respectively.  In comparison, the standard extraction method reaches these
levels for 60\%, 45\% and 15\% of the observations.

\subsection{Typing efficiency}

As mentioned earlier, the most important validation result in the
context of the SEDm is the efficiency of \hypergal to spectrally
classify the target SN.  The test on the simulated cubes is performed
using the same classifier as in ZTF, i.e. \pkg{SNID}; the confidence
criteria given for the classification are however slightly stricter,
as we regularly identified false positives (i.e. SN erroneously
classified as Ia) in the current \pkg{pysedm} pipeline.  The minimum
r$lap$ is set to r$lap_{\min} = 6$ (rather than 5) for the best-fit
model; furthermore, at least 50\% of the top-10 models have to be of
the same type as the best one to confirm a classification.  If one of
these criteria is not met, the spectrum is classified as
``uncertain''.

\begin{figure*}
  \centering
  \includegraphics[width=\linewidth]{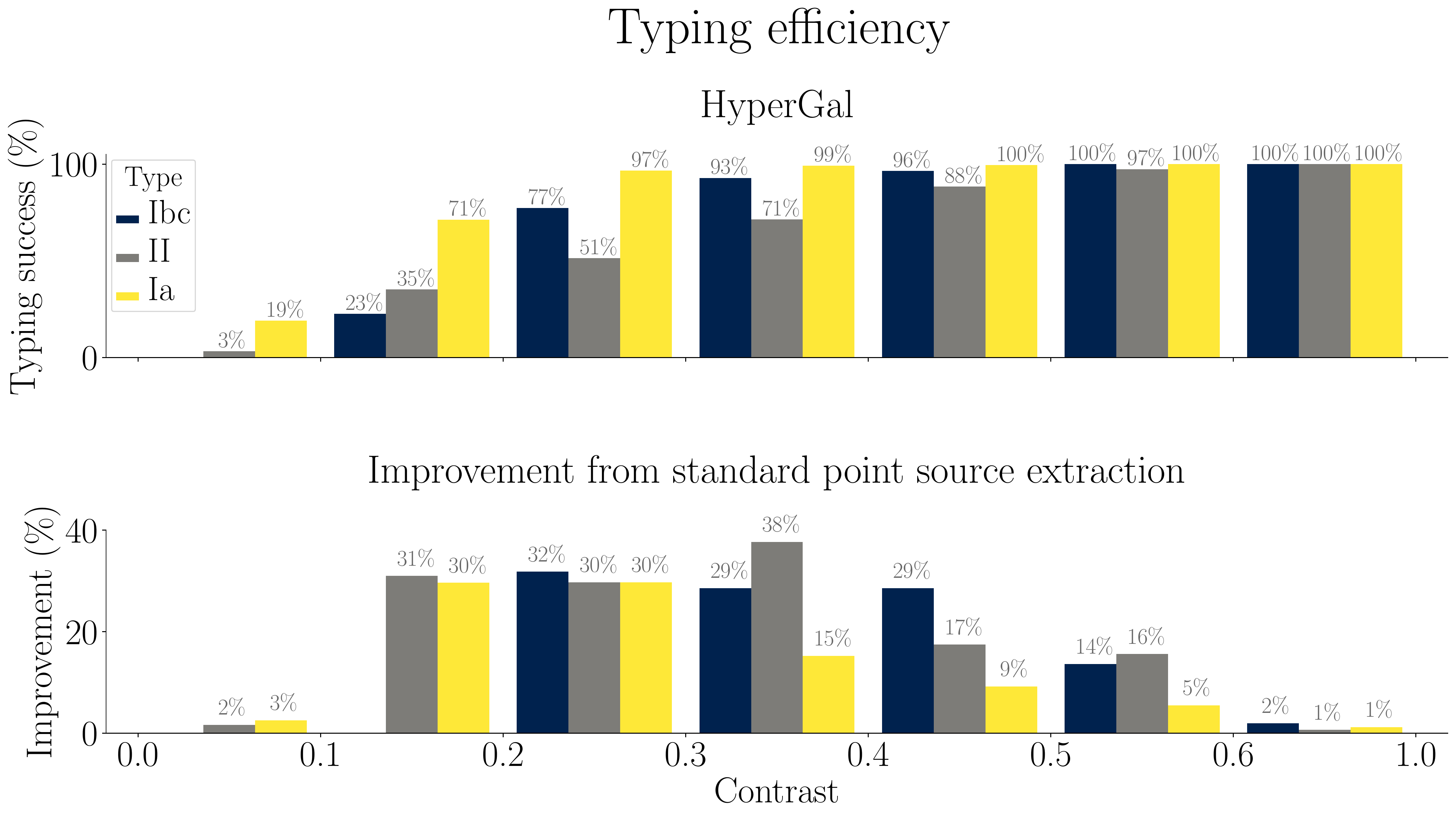}
  \caption{Typing efficiency on the validation simulations.  \emph{Top
      panel}: rate of successful classification with \hypergal for
    each type of SN at different contrast levels.  Results for
    $c > 0.6$ are aggregated as the results vary very little.
    \emph{Bottom panel}: improvement in typing compared to the
    standard extraction method.}
  \label{fig:typingresult}
\end{figure*}

Figure~\ref{fig:typingresult} shows the typing efficiency from
\hypergal, and the improvement with respect to the standard extraction
method without host modeling.  Contrary to the previous RMS analysis,
results are presented for each SN type, since the spectral signatures
are different in all SN types.

As anticipated in Sec.~\ref{ssec:rmscontrast}, both methods are
definitely not reliable for contrasts below 0.1.  SNe~Ia are more
easily classified, due to the quantity and strength of features in
their spectra: the typing success is 71\% for SNe~Ia for
$0.1 \leq c \leq 0.2$ ($\sim 7\%$ of real observations); types~Ibc
and~II on the other hand are correctly classified with a success rate
of 23\% and 35\% respectively.

For $0.2 < c < 0.3$, the typing success reaches more than 96\% for
SNe~Ia, 77\% for~Ibc and 51\% for SNe~II.  More than 99\% of SNe~Ia
are correctly classified with $c > 0.3$, and more than 95\% of all SNe
for $c > 0.4$.  With $\sim 84\%$ of observations having a contrast
$c > 0.3$, $\sim 9\%$ with $0.2 < c < 0.3$ and $\sim 7\%$ with
$0.1 < c < 0.2$, one can conclude that \hypergal can successfully
classify nearly 95\% of all SNe~Ia observed by SEDm.  For a contrast
$c \gtrsim 0.2$ (which represents more than 90\% of the real
observations), nearly 99\% of SNe~Ia are properly classified.  The
improvement brought by \hypergal over the standard extraction method
is obvious, with a sweet spot in $0.1 < c < 0.6$: this will results in
more than 30\% of additional SNe correctly classified.

The main spectral feature of SNe~II being the H$\alpha$ emission line,
usually highly contaminated by the host galaxy, \hypergal allows a
significant improvement for this particular type, from 15\% to 37\% of
additional correctly classified SNe~II in the $0.1 < c < 0.6$ range;
for SNe~Ibc, the difference only appears from $c > 0.2$, with similar
gains between 13\% and 31\%.  SNe~Ia having a lot of strong and easily
identified spectral features, the boost from the standard method is
slightly less manifest, but stays highly significant, from 30\% of
additional correctly classified SNe~Ia for $0.1 < c < 0.2$ to 5\% when
$0.5 < c < 0.6$.  For $c > 0.6$, when the SN ostensibly stands out of
the galaxy, the difference between the two methods becomes marginal
whatever the SN type.

Taking into account the contrast distribution of the observations,
\hypergal should significantly improve the classification of SNe~Ia in
nearly 50\% of the observations (the other half being also properly
classified by the standard extraction method).  As 50\% of the
observations have $0.1 < c < 0.6$, \hypergal will allow the correct
classification of almost 20\% more SNe~Ia in this interval,
corresponding to 10\% of all SNe~Ia classifiable with the SEDm.
Assuming a similar contrast distribution for all SN types, \hypergal
will classifiy 14\% additional SNe~II and 11\% SNe~Ibc.

\begin{figure}
  \centering
  \includegraphics[width=\linewidth]{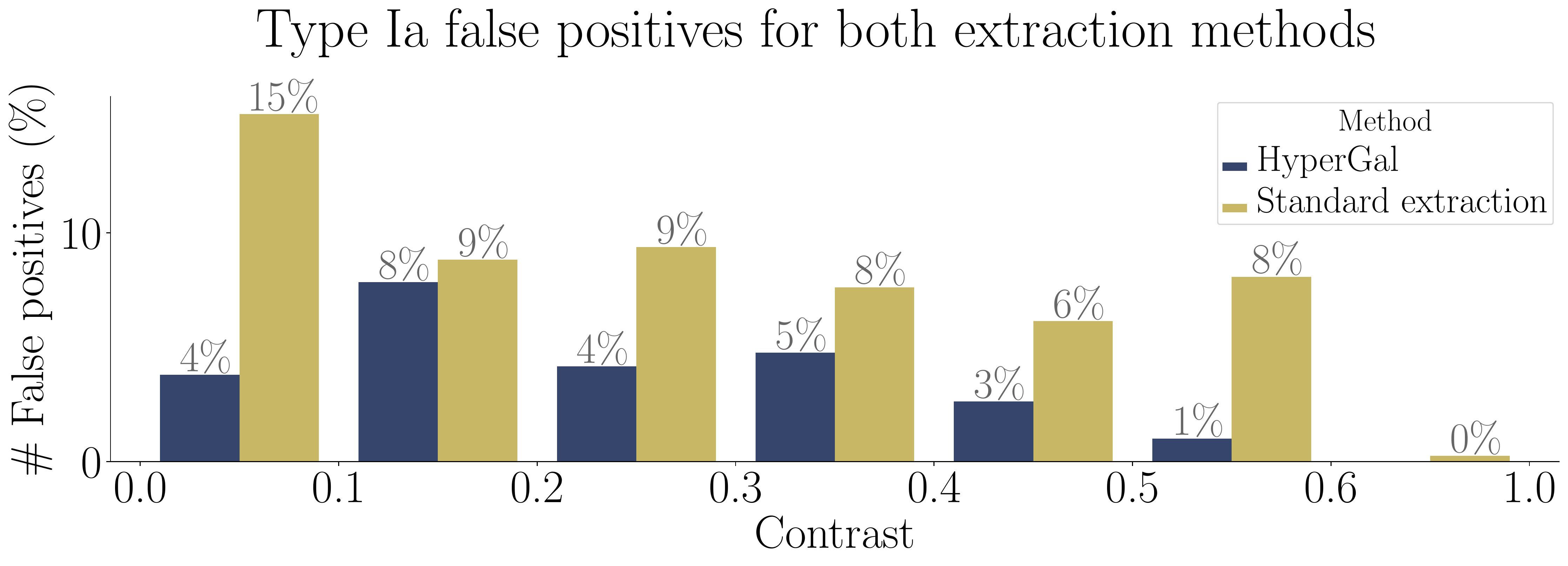}
  \caption{False-Positive Rate in SN~Ia classification for both
    extraction methods as a function of contrast.}
  \label{fig:falspositive}
\end{figure}

To probe the critical contamination of the SN~Ia sample by
core-collapse SNe, the False Positive Rate (FPR) for SN~Ia is
examined.  Figure~\ref{fig:falspositive} shows that \hypergal has a
significantly lower FPR than for the standard method.  Excluding the
unrealistically low contrast cases ($c < 0.1$), \hypergal shows a
progressive decrease in FPR from 8\% to 1\% for contrast rising from
0.1 to 0.6 (FPR is null beyond that); in comparison, the standard
method oscillates between 6 and 9\% in same contrast range.  As a
conclusion, the \hypergal FPR is on average less than 5\% for
contrasts between 0.1 and 0.6 ($\sim 50\%$ of the observations), and
less than 2\% for $c > 0.1$ (more than 99\% of all observations); this
is half as much as the standard extraction method.

\section{Discussion}
\label{sec:discuss}

We now discuss some limitations of the current \hypergal
implementation, and possible future developments.

Regarding the validation methodology, we acknowledge some
simplifications with respect to actual observations.  For instance,
the true distance distribution between the SN and its host was not
explicitly modeled, i.e. this parameter was marginalized uniformly
between 0 and $5 \farcs 6$.  As a full-scene modeler which properly
handles this parameter and therefore shows little sensitivity to it
(Sec.~\ref{ssec:rmscontrast}), this approximation does not impact the
\hypergal results; this is not true for the single point-source method
which critically depends on the transient-host distance.  Overall, we
think the validation approximations actually tend to minimize the
improvement of \hypergal with respect to the standard method.

Undoubtedly, the most limiting constraint from \hypergal is the need
for an external redshift measurement of the host galaxy, an a priori
needed by the SED fitter used as a physically-motivated host galaxy
spectral interpolator, and of critical importance for the treatment of
emission lines.  In practice, this is not so much of an issue: in the
current ZTF sample, about 50\% of SN hosts already have a spectral
redshift, mostly from SDSS surveys \citep{ztfspecred}, with a
precision of $\sigma_z\sim\num{e-5}$ for $z<0.1$ \citep{Bolton2012};
the remaining 50\% of SNe have a redshift deduced from a preliminary
extraction of the SN spectrum, either from low-resolution spectral
features in the SN spectrum ($\sim 40\%$) or emission lines of the
host galaxy having contaminated the SN spectrum ($\sim 10\%$).  In
both cases, the redshift is estimated by \pkg{SNID} with a precision
of $\sigma_z\sim\num{5e-3}$ \citep{ztfspecred}.  Furthermore, 95\% of
ZTF SN hosts are brighter than 20~mag, allowing other surveys such as
the Dark Energy Spectroscopic Instrument (DESI) Bright Galaxy Survey
\citep{DESI} to systematically provide a large fraction of spectral
redshifts in the future.

A slightly incorrect input redshift (encoded as a wavelength offset of
the emission line position in the hyperspectral galaxy model), as well
as an approximate SED fit of the emission line fluxes (marginally
constrained by broadband photometric observations) is
corrected to
first order by the monochromatic galaxy amplitudes $G(\lambda)$ during
the ultimate 3D fit.  Primarily introduced to recover flux calibration
mismatch between PS1 and SEDm, this normalization parameter actually
interferes in a non-trivial way with the position and intensity of
emission lines in the brightest parts of the scene to minimize
residuals between fixed (at this stage of the procedure) hyperspectral
model and SEDm observations.  This particular effect, which depends on
the relative distribution of stellar and gaseous components in the
host, has not been studied extensively for \hypergal, but we note it
is efficient to disentangle host spectral features from SN spectrum
even with sub-optimal input redshift and/or emission line fluxes.
However, it effectively precludes the use of the residual host
component for any \emph{a posteriori} measurements, e.g. redshift or
local measurement of H$\alpha$ flux, yet crucial for local environment
studies mentioned earlier \citep[e.g.][]{lssf_rigault20}.

One could think of including a consistent redshift estimate directly
in the \hypergal procedure, at the level of the hyperspectral model
(to minimize artificial fluctuations of $G(\lambda)$), but also at the
level of the SN spectral typing (to reach a redshift consensus between
the host and the SN).  This would imply to include the intensive SED
fit and/or the SN typing procedure in the minimization loop,
computationally costly in either cases.  Another major \hypergal
development would be to use the SEDm cube, a rich and faithful
observation of the host galaxy at the position of the transient, as
additional hyperspectral constraints in the SED fitting process.  Both
developments would push the concept of an SED fitter merely used as a
spectral interpolator to its limit.  It would then probably be
preferable to switch to other more efficient methods such as
physics-enabled deep learning \citep{2021AJ....162..275B}.

\section{Conclusion}

This paper presents \hypergal, a fully automated scene modeler for the
transient typing with the SEDm \citep{SEDm}.  The core of this
pipeline is based on the use of archival photometric observations of
the host galaxy, taken before the SN explosion.  Knowing the physical
processes in place within galaxies, as encoded in the SED fitter
\pkg{cigale}, the spectral properties of the host are modeled,
adjusted, and scaled appropriately to create a hyperspectral model of
the host galaxy.  This 3D intrinsic model is then convolved with the
spectro-spatial instrumental responses of the SEDm, and projected in
the space of the observations.  A full scene model, including the
structured host galaxy, the point source transient and a smooth
background, is finally produced to match the SEDm observations,
allowing the extraction of the SN spectrum from a highly contaminated
environment.

The pipeline is validated on a large set of realistic simulated SEDm
observations, covering a wide variety of observation conditions
(airmass, seeing and PSF parameters), scene details (host morphology,
distance to the host, host/SN contrast) and transient types.  The
contrast distribution is estimated from about 3000 observed SNe~Ia of
the ZTF Cosmology SN~Ia DR2 paper to come \citep{dr2rigault}.  The
transient spectra in the 5000 simulations are then extracted with
\hypergal and compared to the historical point-source method, which
ignores the structured host component.

The most important results concern \hypergal efficiency in
spectroscopically typing SNe, a key objective of the SEDm instrument.
The full scene modeler shows an ability to correctly classify
$\sim 95\%$ of the observed SNe~Ia under a realistic contrast
distribution.  For a contrast $c \gtrsim 0.2$ (more than 90\% of the
observations), nearly 99\% of the SNe~Ia are correctly classified.
Compared to the standard extraction method, \hypergal correctly
classifies nearly 20\% more SNe~Ia between $0.1 < c < 0.6$,
representing $\sim 50\%$ of the observation conditions.

The false positive rate for \hypergal is less than 5\% for contrasts
between 0.1 and 0.6, and less than 2\% for $c > 0.1$ ($> 99\%$ of the
observations); this is half as much as the standard extraction method.

\hypergal has demonstrated its ability to extract and classify the
spectrum of an SN even in the presence of strong contamination from
its host galaxy.  The improvement compared to the standard method is
significant: this will noticeably improve the statistic of the SNe~Ia
sample for the ZTF survey while reducing a potential environmental
bias, and will ultimately impact the precision of the cosmological
analyses.

\begin{acknowledgements}
  This project has received funding from the Project IDEX-LYON at the
  University of Lyon under the Investments for the Future Program
  (ANR-16-IDEX-0005), and from the
  European Research Council (ERC) under the European Union's Horizon
  2020 research and innovation programme (grant agreement n°759194 -
  USNAC). The SED Machine is based upon work supported by
  the National Science Foundation under Grant No. 1106171. Based on
  observations obtained with the Samuel Oschin Telescope 48-inch and
  the 60-inch Telescope at the Palomar Observatory as part of the
  Zwicky Transient Facility project. ZTF is supported by the National
  Science Foundation under Grant No. AST-1440341 and a collaboration
  including Caltech, IPAC, the Weiz- mann Institute for Science, the
  Oskar Klein Center at Stockholm University, the University of
  Maryland, the University of Washington, Deutsches Elektronen-
  Synchrotron and Humboldt University, Los Alamos National
  Laboratories, the TANGO Consortium of Taiwan, the University of
  Wisconsin at Milwaukee, and Lawrence Berkeley National
  Laboratories. Operations are conducted by COO, IPAC, and UW.

  This research made use of \pkg{python} \citep{python3},
  \pkg{astropy} \citep{Astropy, Astropy2}, \pkg{matplotlib}
  \citep{matplotlib}, \pkg{numpy} \citep{numpy1, numpy2}, \pkg{scipy}
  \citep{scipy1, scipy2}.  We thank their developers for maintaining
  them and making them freely available.
\end{acknowledgements}

\bibliographystyle{aa}
\bibliography{reference}

\end{document}